\documentclass[lettersize,journal]{IEEEtran}
\usepackage{amsmath,amsfonts}
\usepackage{algorithmic}
\usepackage{algorithm}
\usepackage{array}
\usepackage[caption=false,font=normalsize,labelfont=sf,textfont=sf]{subfig}
\usepackage{textcomp}
\usepackage{stfloats}
\usepackage{url}
\usepackage{verbatim}
\usepackage{graphicx}
\usepackage{array}
\usepackage{cite}
\usepackage{hyperref}
\hypersetup{hidelinks,
	colorlinks=true,
	allcolors=red,
	pdfstartview=Fit,
	breaklinks=false}
\begin{document}

\title{Development of Skip Connection in Deep Neural Networks for Computer Vision and Medical Image Analysis: A Survey}

\author{Guoping Xu, Xiaxia Wang, Xinglong Wu, Xuesong Leng, Yongchao Xu
\thanks{Guoping Xu, Xiaxia Wang, Xinglong Wu, Xuesong Leng are with the School of Computer Science and Engineering, Hubei Key Laboratory of Intelligent Robot, Wuhan Institute of Technology, Wuhan, Hubei, China, 430205.}
\thanks{ Yongchao Xu is with the School of Computer Science, Wuhan University, Wuhan, Hubei, China, 430072.}
\thanks{ E-mail:xugp@wit.edu.cn, wxxvva@163.com, xwu@wit.edu.cn, 2302448483@qq.com, yongchao.xu@whu.edu.cn}
\thanks{ (Corresponding author: Xinglong Wu)}}
\markboth{Journal of \LaTeX\ Class Files,~Vol.~14, No.~8, August~2021}%
{Shell \MakeLowercase{\textit{et al.}}: A Sample Article Using IEEEtran.cls for IEEE Journals}


\maketitle

\begin{abstract}
Deep learning has made significant progress in computer vision, specifically in image classification, object detection, and semantic segmentation. The skip connection has played an essential role in the architecture of deep neural networks, enabling easier optimization through residual learning during the training stage and improving accuracy during testing. Many neural networks have inherited the idea of residual learning with skip connections for various tasks, and it has been the standard choice for designing neural networks. This survey provides a comprehensive summary and outlook on the development of skip connections in deep neural networks. The short history of skip connections is outlined, and the development of residual learning in deep neural networks is surveyed. The effectiveness of skip connections in the training and testing stages is summarized, and future directions for using skip connections in residual learning are discussed.  Finally, we summarize seminal papers, source code, models, and datasets that utilize skip connections in computer vision, including image classification, object detection, semantic segmentation, and image reconstruction. We hope this survey could inspire peer researchers in the community to develop further skip connections in various forms and tasks and the theory of residual learning in deep neural networks. The project page can be found at \href{https://github.com/apple1986/Residual_Learning_For_Images}{ResidualLearningSurvey}

\end{abstract}


\section{Introduction}
\label{sec1}
\IEEEPARstart{A}{t} present, deep neural networks have led to a series of breakthroughs in computer vision \cite{he2016deep}, natural language processing \cite{vaswani2017attention}, and speech recognition \cite{graves2012long}. However, it has gone through nearly a century of development. In the early stage, inspired by the connections of neurons in the human brain,  McCulloch and Pitts proposed the first artificial neuron model in 1943 \cite{mcculloch1943logical}, MP (McCulloch-Pitts Neuron). The MP treats neural events and the relations between them as propositional logic. Regrettably, it contained no learnable parameters and could not be learned from input data. In 1949, the psychologist Hebb \cite{attneave1950organization} proposed a learning rule. It said that if two neurons have outputs x and y, and x excites (fires) y, the connection's strength increases, and vice versa. Inherited from MP and the Hebbian learning rules, the perceptron was introduced by Rosenblatt in 1958 \cite{rosenblatt1958perceptron}. It is a learnable and linearly separable classifier that is a single-layer neural network using a step function as the activation function. In 1962, Hubel and Wiesel found the specific patterns that stimulated activity in specific parts of the  visual cortex of the cat \cite{hubel1962receptive}. Afterward, Fukushima proposed a neural network model, Neocognitron \cite{fukushima1979neural}\cite{fukushima1980neocognitron}, imitated from Hubel and Wiesel’s hierarchy model for the visual nervous system. The Neocognitron is a self-organizing architecture and could learn stimulus patterns unaffected by position shift. Yet, owing to the lack of a proper algorithm for training the model to update parameters automatically, its application is limited. In 1981, Werbos introduced error backpropagation (BP) into neural networks \cite{werbos2005applications}. 
Later, several works were published in subsequent years using backpropagation (BP) in neural networks\cite{lecun1988theoretical}. In particular, in 1986, Rumelhart et al. discovered that the hidden layers of a neural network could learn internal representations with BP, which led to the algorithm's popularity\cite{rumerhart1986learning}. In 1988, LeCun et al. presented the first trainable convolutional neural network, inspired by the structure of the Neocognitron and BP \cite{lecun1989backpropagation}. Convolutional neural networks use fixed-size filters, known as receptive fields, to perform convolution operations. This contrasts with fully connected neural networks, where all output neurons are connected to all input neurons. When examining the history of neural networks, it becomes clear that the architecture and learning algorithm are closely intertwined. 

Many comprehensive survey papers have been published on neural networks and deep learning \cite{srinivas2015deep}\cite{lecun2015deep}, which provide a thorough introduction to their development. Some survey papers focus on the progress of deep learning in specific topics, such as image and text classification \cite{wang2019development}\cite{minaee2021deep}, object detection \cite{oksuz2020imbalance}\cite{zou2023object}, loss functions \cite{ma2021loss}\cite{jadon2020survey}\cite{wu2022iou}, semantic segmentation \cite{liu2021review}\cite{asgari2021deep}\cite{minaee2021image}, optimization methods \cite{sun2020optimization}\cite{feng2018overview}\cite{le2011optimization}, regularization \cite{kukavcka2017regularization}\cite{goodfellow2016regularization}\cite{neyshabur2017implicit}, activation functions \cite{szandala2021review}\cite{apicella2021survey} etc. However, one of the most popular "tricks" in the design of deep neural networks, skip connections or residual learning, has not yet been thoroughly discussed in these surveys. These techniques are commonly utilized to design the architecture of deep neural networks because they accelerate training speed and improve final performance. In this paper, we survey the development of skip connections in deep neural networks for computer vision, discuss their effectiveness, and explore future directions for using skip connections in designing architectures of deep neural networks. To the best of our knowledge, it is the first survey to discuss the skip connection for residual learning in deep neural networks.

The rest of this survey is structured as follows: Section \ref{sec1} provides a brief history of residual representation and skip connections. In Section \ref{sec2}, we will examine the origin of skip connections in neural networks. Section \ref{sec3} lists important backbone architectures that inherit the idea of residual learning. Section \ref{sec4} discusses the theoretical exploration of why skip connections effectively train deep neural networks. Finally,  the left sections  present some open questions, public resources, and a summary of this paper.

\section{The Origin of Skip Connection and Residual Learning}
\label{sec2}
\begin{figure}[!t]
\centering
\includegraphics[width=2.5in]{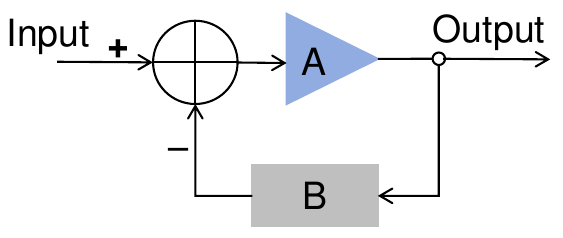}
\caption{A negative feedback system. Block A processes the signal, while Block B responds to bring the system back to a stable state with the feedback signal.}
\label{fig1}
\end{figure}

Skip connection, also known as shortcut connection, has been studied for a long time. In 1948, Wiener introduced negative feedback into the control system and proposed Cybernetics \cite{wiener2019cybernetics}. Negative feedback refers to the output of a system being fed back to the input to promote the system’s stability. Fig. \ref{fig1} illustrates the basic concept of negative feedback, in which the output signal can adjust the input through feedback.

Well-known RNN architectures, such as the Hopfield network \cite{hopfield1982neural}, Long Short-Term Memory (LSTM) \cite{hochreiter1997long}, and Gated Recurrent Unit (GRU) \cite{dey2017gate}, aim to store information learned from the previous input sequences. Fig. 3 demonstrates one typical unit of LSTM. Thanks to its gate mechanism, the LSTM can effectively pass historical information to the current input. When comparing the negative feedback system in Fig. 1 and LSTM in Fig. 3, both can provide feedback to the input of a control system or the hidden layer of LSTM through some adjustable circuits or learnable gates. If all these gates are “open,” such feedback degrades into skip connections. 

\begin{figure}[!t]
\centering
\includegraphics[width=3.5in]{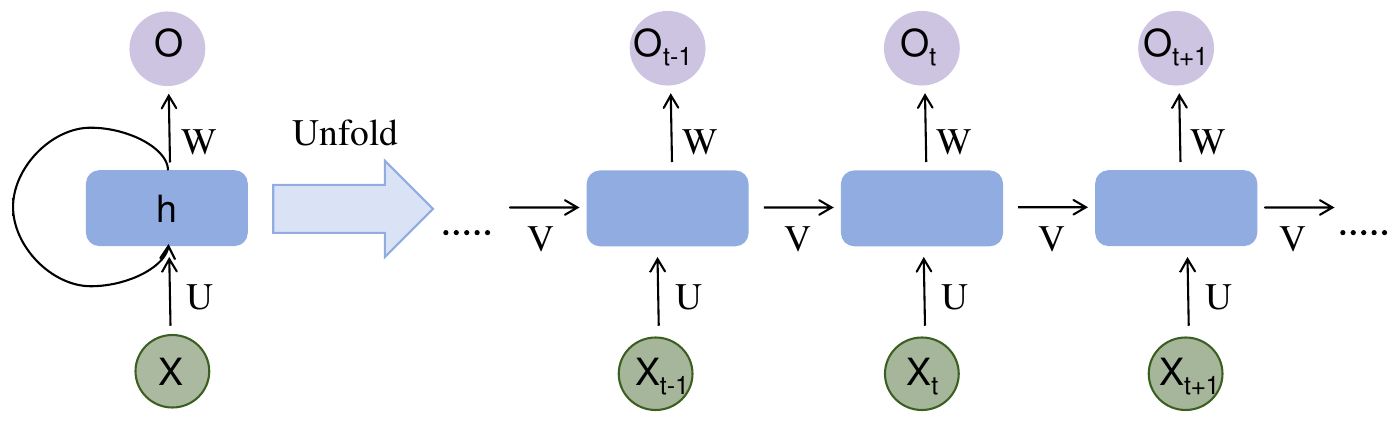}
\caption{A unit of RNN and the unfolded form. }
\label{fig2}
\end{figure}

\begin{figure}[!t]
\centering
\includegraphics[width=3.5in]{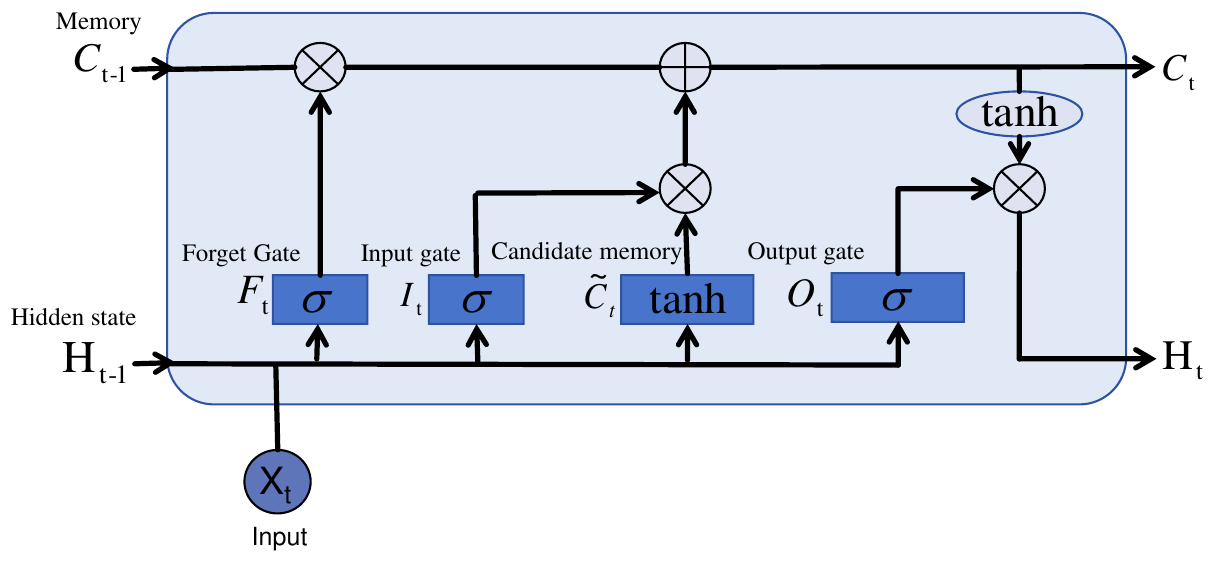}
\caption{The architecture of an LSTM cell. It includes the forget gate, input gate, and output gate. These gates control the flow of information into and out of the LSTM cell.}
\label{fig3}
\end{figure}

The concept of residual representations is also adopted in image processing and optimization. Similar to the multiresolution signal decomposition based on the wavelet representation, the residual signal is represented by a wavelet orthonormal basis \cite{mallat1989theory}. In \cite{burt1987laplacian}, a compact method for image coding called Laplacian pyramid was proposed. Compared to the Gaussian pyramid, in which subsequent images are weighted down using a Gaussian average (Gaussian blur) and scaled down, the Laplacian pyramid keeps the residual image of the blurred version at each level. This method is efficient for image denoising and reconstruction. Different from the residual representation of the image directly, the VLAD \cite{jegou2011aggregating} and Fisher Vector encode the residual features (vectors) into a dictionary for image recognition. In \cite{jegou2010product}, it is demonstrated that encoding residual vectors is more effective than encoding original vectors. Previous studies have shown that the idea of skip connections or residual representations is an effective way to process images and represent features.

\section{Skip Connection in Residual Neural Network}
\label{sec3}
Deep convolutional neural networks have made significant progress in image classification since the ImageNet ILSVRC (Large scale visual recognition challenge)-2010 and ILSVRC-2012 contests \cite{russakovsky2015imagenet}. In 2012, Krizhevsky, Sutskever, and Hinton proposed a deep convolutional neural network, which was later named AlexNet \cite{krizhevsky2012imagenet}. The superior performance of AlexNet for image classification has generated interest in academia and industry. The idea of “deep” neural networks became widely accepted and was utilized in the top results of the ILSVRC from  2012 to 2014 \cite{zeiler2014visualizing}\cite{simonyan2014very}\cite{szegedy2015going}. As Bengio et al. claimed, “deep” networks offer advantages in both computation and statistics for complex tasks \cite{bengio2013representation}.

However, increasing the number of layers does not necessarily improve the accuracy of neural networks. Accuracy may reach a saturation point and then rapidly decrease \cite{srivastava2015highway}\cite{he2015convolutional}. Therefore, increasing the number of hidden layers is insufficient solely to make neural networks “deep.” One reason for this is that training in very deep networks can be challenging  to optimize due to the risk of vanishing or exploding gradients \cite{hochreiter1997long}. There are some techniques to aid in the  training of deep networks, such as initialization strategies \cite{glorot2010understanding}\cite{saxe2013exact}\cite{he2015delving}, feature map normalization \cite{ioffe2015batch}\cite{ba2016layer}\cite{wu2018group}, training networks in multiple stages \cite{simonyan2014very}\cite{srivastava2015training}, or designing multiple loss functions to specific layers \cite{szegedy2015going}\cite{lee2015deeply}. These methods can alleviate the problem of vanishing gradients. However, training networks with more than 50 layers were challenging before 2015.

Inspired by the gating mechanism and, in particular, the success of LSTM, Highway networks \cite{srivastava2015highway} were proposed to facilitate gradient-based training of “deep” neural networks. This architecture relies on specially designed gating units, which are used to regulate the transmission of feature maps through a network. The gating mechanism allows feature maps to flow through multiple layers without saturation in Highway networks. Compared to the negative feedback of control theory or Cybernetics for system stability \cite{wiener2019cybernetics}, the gating mechanism in the feedforward Highway network plays as the flow of modulating signals for feature transmission. Since the gating units are data-dependent and have learnable parameters, it might have a negative impact on the information (feature maps) flowing during training. For instance, if the gating units do not modulate the feature maps well, some discriminative representation could be prevented from passing through the subsequent layers. In other words, the gating units should fit the parallel branch well during training, which increases the difficulty of achieving superior performance \cite{he2016identity}.

Unlike the gate-mechanism methods, where the information flow is modulated by the learnable gates, ResNet was proposed to integrate solely skip connections to the deep neural networks \cite{he2016deep}. Formally, the ResNet is a particular case where all the gates of Highway Networks are open, but it is difficult to achieve because each gate contains independent learnable parameters. It is almost impossible to make all these parameters become zero when training a Highway Network. More importantly, compared to the Highway networks, the primary motivation of the ResNet is different. In ResNet, the layers are reformulated as learning residual functions with reference to the inputs by introducing a skip connection. Compared to learning the unreferenced feature maps, it becomes easier to optimize the residual features and achieve continuous gain with increasing depth. The main block of the Highway network and the ResNet are illustrated in Fig. 4.

\begin{figure}[http]
\centering
\includegraphics[width=3.5in]{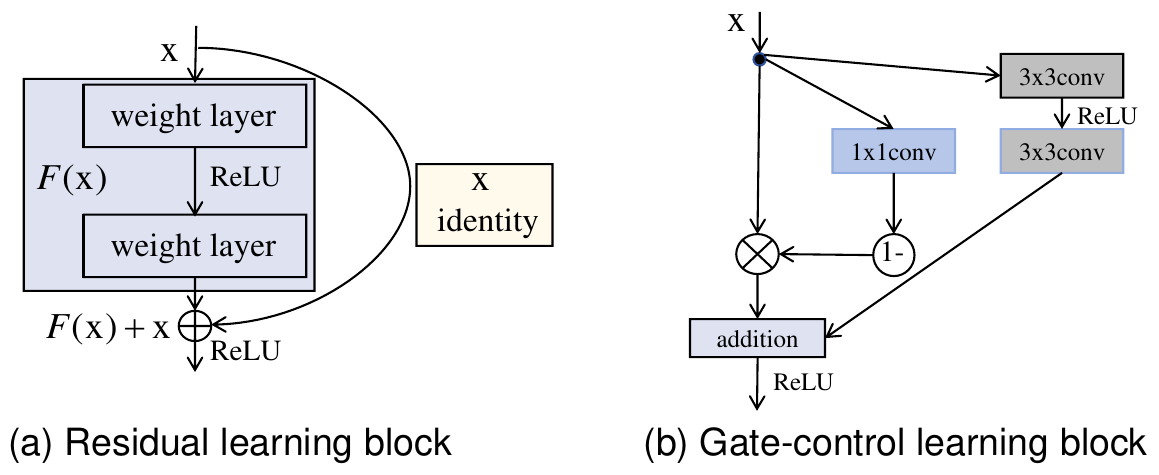}
\caption{Residual learning block and gate learning block. In (a), the residual leaning block consists of a residual unit (with light blue box) and an identity map with a skip connection (light yellow). In (b), the input x is modulated by a 1x1 convolution block. Meanwhile, the input x is also processed by two convolutional operations for feature learning. The output is the sum of the modulated input x and the output of the feature learning branch.}
\label{fig4}
\end{figure}

In Fig. \ref{fig4}, $F(x)$ denotes the residual unit, and $x$ represents the identity map realized by a short connection. The output $F(x) + x$ are the final feature maps (omitting the ReLU function). In this paper, we refer to the block as the residual block. Compared to the previous simple stacking layers, the ResNets can address the degradation problem well in training a deep neural network when deeper networks are able to start converging, and the accuracy gets saturated. In other words, it makes the “deep” neural networks possible. Due to its high performance in image classification, localization, detection, and segmentation, the residual block has become the basic default setting in many follow-on architectures, and the ResNets have been the baseline for comparison in many computer vision tasks.

\begin{figure}[http]
\centering
\includegraphics[width=3.5in]{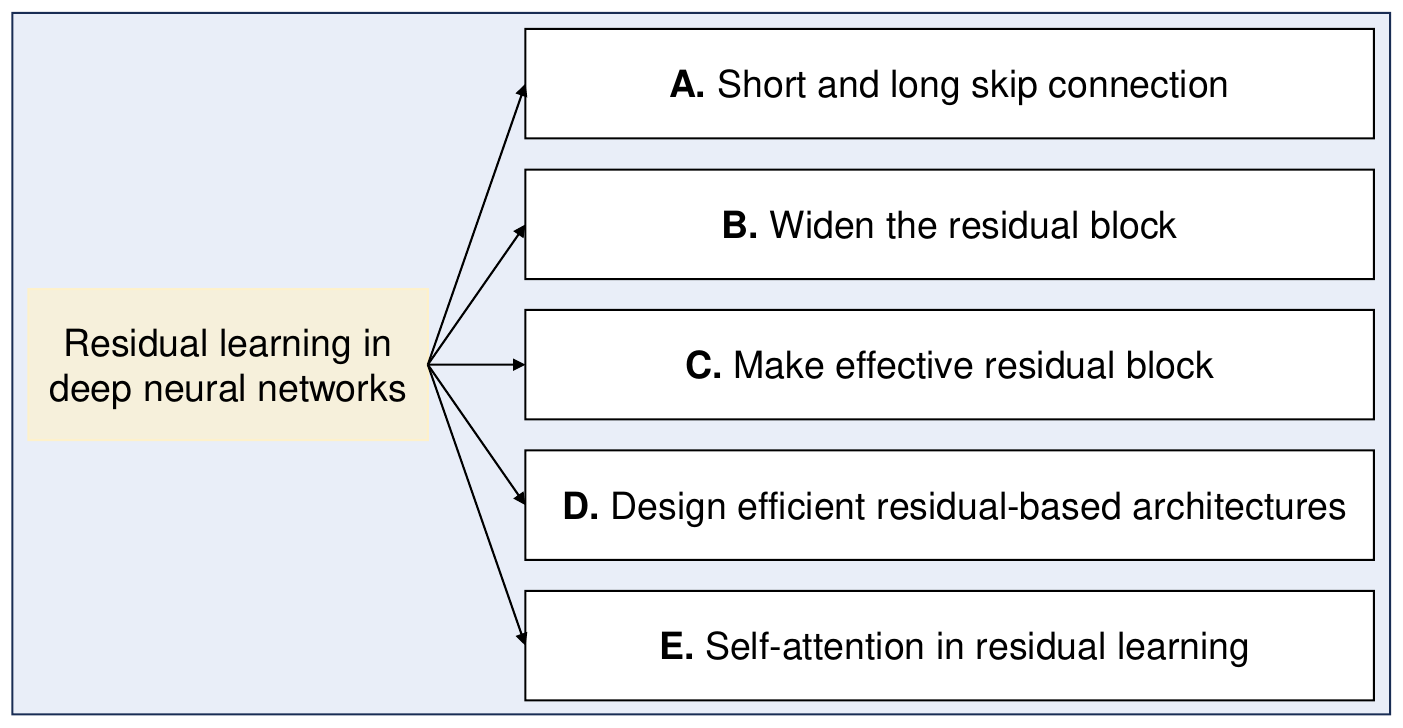}
\caption{An overview of the evolution of the skip connection (residual learning). Note that the skip connection and residual learning have the same meaning. The skip connection in deep neural networks is used for residual learning.}
\label{fig5}
\end{figure}

\section{Development of skip connections (residual learning) in deep neural network}
\label{sec4}
Although the ResNets achieved significant performance compared to previous methods on the ImageNet and COCO datasets for the task of image classification, detection, and segmentation in 2015, some issues also deserve to be exploited. In this section, we categorize these problems into five groups. In Section A, we discuss the length of the skip connection for residual learning. The effectiveness of widening the residual block was presented in Section B. In Sections C and D, we surveyed seminal approaches for improving the accuracy or efficiency of residual learning. Finally, we investigate the attention mechanism for residual learning. The main contents are summarized in Fig. \ref{fig5}.

\subsection{From short skip connection to long skip connection}
The original residual block consists of only two convolution layers and a relatively short skip connection. The question arises as to whether longer skip connections would be beneficial. Specifically, what would happen if a neural network contained both long and short skip connections? For this discussion, a long skip connection is defined as one that crosses at least three convolution layers.

\begin{figure}[http]
\centering
\includegraphics[width=3.5in]{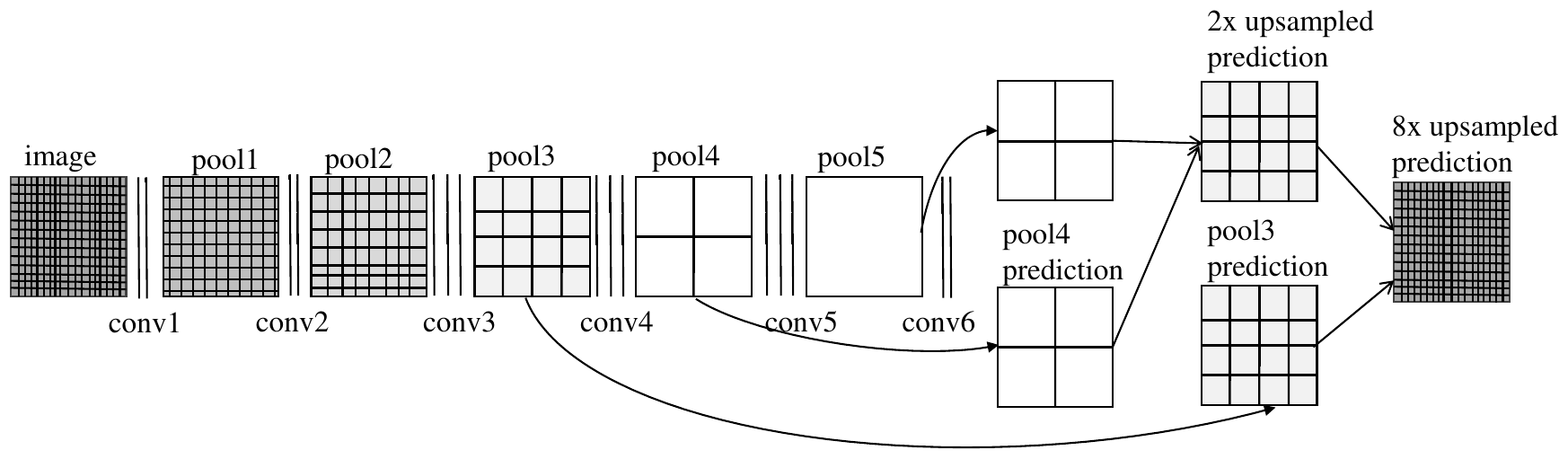}
\caption{Architecture of the fully convolutional neural network (FCN-8s). Here, two skip connections are utilized to pass the previous feature maps and concatenate them with the last output of the convolution layer. Finally, the multi-scale feature maps are upsampled to the same spatial size and concatenated for semantic segmentation. }
\label{fig6}
\end{figure}

\subsubsection*{\bf Long skip connections}
In \cite{long2015fully}, the authors proposed a fully convolutional network (FCN) for semantic segmentation. The FCN can take into input images of arbitrary size and produce corresponding pixel-wise segmentation output by replacing the last fully connected layer with a convolutional layer. Then, the long skip connections are used to combine detailed information from shallow layers with semantic information from deep layers. This aims to produce more accurate representative features for semantic segmentation (Fig. \ref{fig6})

In 2015, U-Net was proposed as another important architecture with skip connections \cite{ronneberger2015u}. It comprises an encoder and a decoder. The encoder extracts semantic information for object recognition through a series of convolution, batch normalization, ReLU, and max-pooling layers. The decoder completes the segmentation task of objects through upsampling, convolution, batch normalization, and ReLU. Similar to FCN, U-Net utilizes long skip connections to combine the local information from the encoder with context information from the decoder. In this case, the concatenated feature maps contain more information, improving the performance of object detection and pixel segmentation. The U-Net has become a widely used benchmark in medical image segmentation due to its simplicity and generality (Fig. 7).

\begin{figure}[http]
\centering
\includegraphics[width=3.5in]{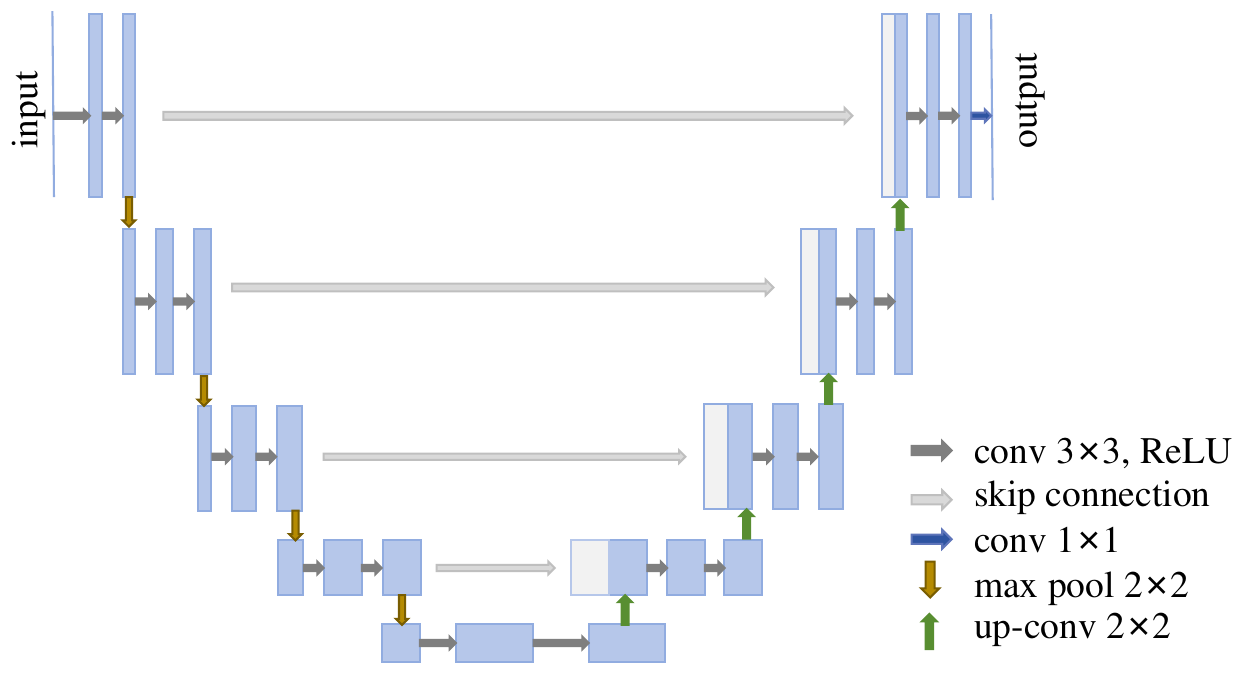}
\caption{ The Architecture of the U-Net network, which utilizes long skip connections to concatenate the feature maps for the corresponding encoder and decoder. }
\label{fig7}
\end{figure}
In addition, there are works that take advantage of long skip connections to combine local and semantic feature maps to enhance segmentation performance, such as SegNet \cite{badrinarayanan2017segnet}, DiSegNet \cite{xu2021disegnet}, UNet++ \cite{zhou2019unet++}, nnU-Net \cite{isensee2021nnu} and RED-Net \cite{mao2016image}. In summary, the long skip connection can pass more feature details from the previous encoder convolutional layers to the decoder convolutional layers. This helps with information aggregation and improves pixel-intensive tasks, such as semantic segmentation, object detection, and image restoration.

\subsubsection*{\bf Integration of short and long skip connections}
The use of short skip connections can help alleviate the degradation problem by making it easier to learn residual feature maps compared to directly fitting a desired underlying feature mapping. Meanwhile, long skip connections bridge the gap between detailed and semantic information, demonstrating their effectiveness in many tasks.

In \cite{huang2017densely}, a type of densely connected convolutional network (DenseNet) was proposed (Fig. 8 ). The DenseNet connects each layer to every other subsequent layer in a feed-forward fashion, including both short and long skip connections. Compared to ResNet, DenseNet makes full use of the feature maps from each layer to ensure maximum information flow between layers in the network. Two of its compelling advantages are strengthened feature propagation and encouraged feature reuse. 

\begin{figure}[http]
\centering
\includegraphics[width=3.5in]{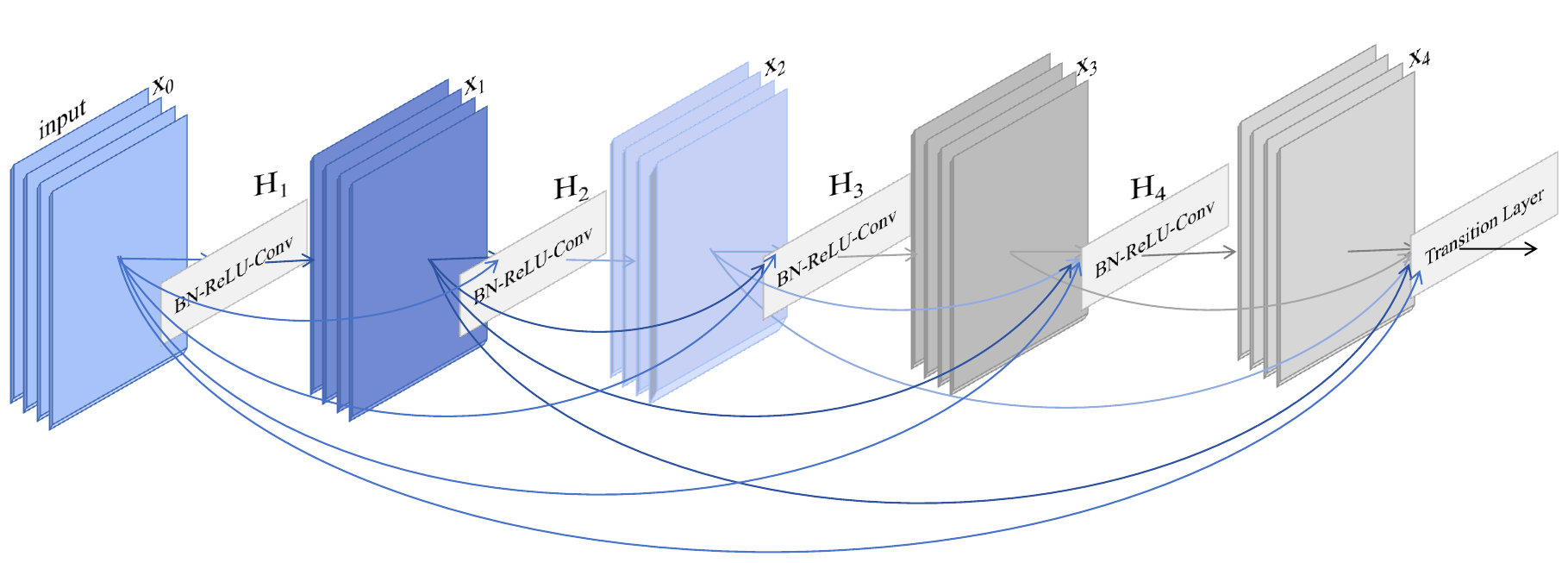}
\caption{ The Architecture of the U-Net network utilizes long skip connections to concatenate the feature maps for the corresponding encoder and decoder. }
\label{fig8}
\end{figure}

Short skip connections are integrated into a 3D U-Net architecture for electric field mapping in transcranial magnetic stimulation \cite{xu2021rapid}. The encoder directly embeds residual blocks, and the decoder receives feature maps from the encoder through long skip connections. Due to the combination of short and long skip connections, the 3D U-Net’s training efficiency and feature learning are improved by enhancing the information aggregation within the U-Net both locally and globally.

Additionally, many works also utilize short and long skip connections for various tasks. For instance, the ResUNet-a \cite{diakogiannis2020resunet} employs a U-Net backbone combined with residual blocks, atrous convolutions, and pyramid scene parsing pooling. The proposal of volumetric ConvNets \cite{yu2017volumetric} included mixed residual connections for prostate segmentation from 3D MR Images. In summary, it is more effective to integrate both short and skip connections for residual feature learning. However, it can be challenging to determine the optimal format to combine them for a specific task.

\subsection{Widen the residual unit}
As the neural networks become deeper with residual blocks, a significant issue arises: training very deep residual networks is time-consuming. In \cite{szegedy2015going}, the GoogLeNet (Inception-v1, as shown in Fig. 9.) was proposed which allows for increasing the depth and width of the network by stacking a series of inception modules. This approach demonstrated the effectiveness of widening the width of networks. (Note that the “wide” means a greater number of feature maps or more branches in a block).
\begin{figure}[http]
\centering
\includegraphics[width=3.5in]{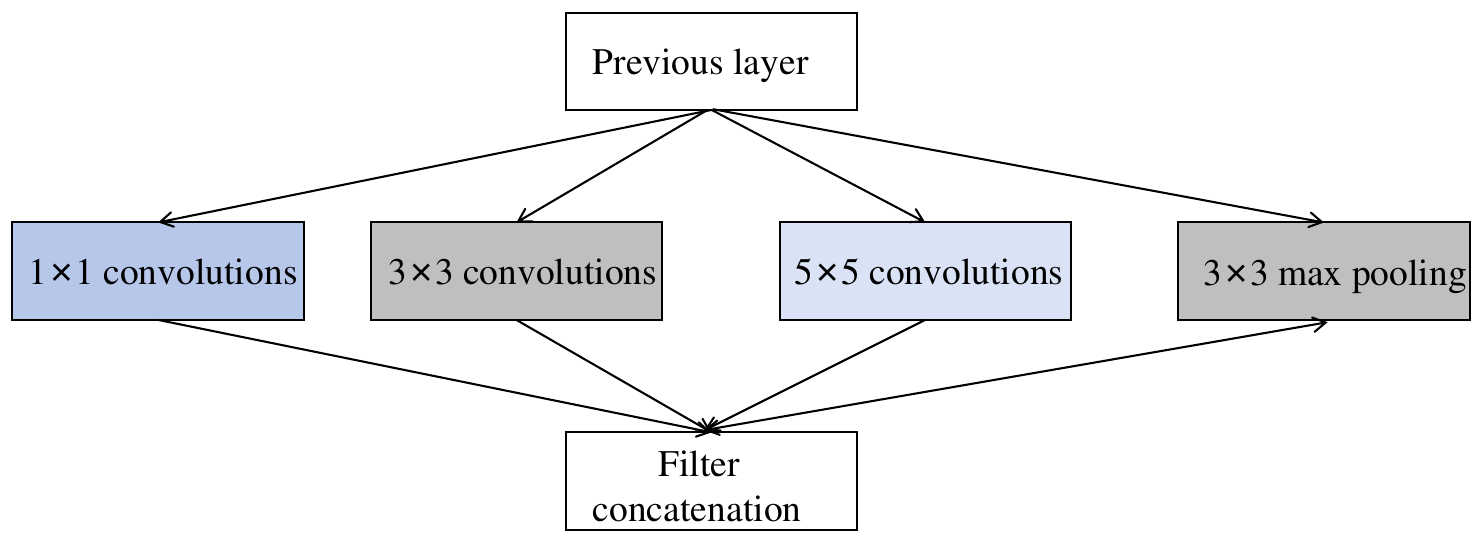}
\caption{Basic Inception module in GoogLeNet}
\label{fig9}
\end{figure}

Due to the effectiveness of skip connections and inception, a natural question is whether are there any benefits to combining the inception module with skip connection. The Inception-v4 \cite{szegedy2017inception} introduced skip connections into the inception module, and the authors found that training with skip connections significantly accelerated the training of Inception networks and achieved better performance compared to similarly expensive Inception networks without skip connections. In simpler terms, incorporating skip connections into the inception module alters the feature learning in a residual manner. Additionally, the widened multi-scale feature maps can run in parallel, making it easier to  optimize residual feature learning.

\begin{figure}[http]
\centering
\includegraphics[width=3.5in]{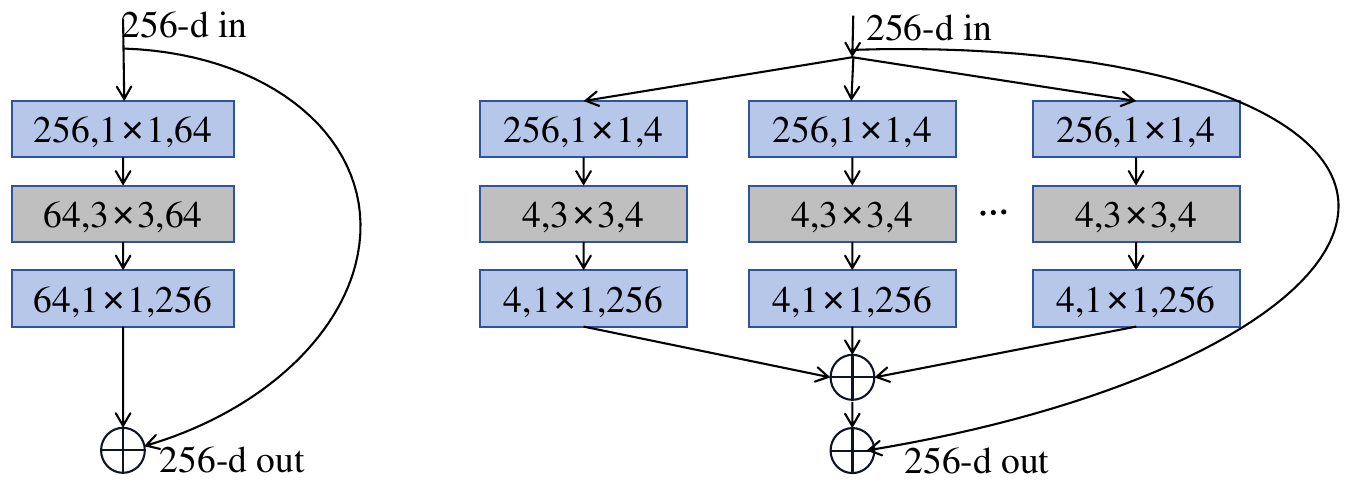}
\caption{A residual learning block (left) and a block from ResNeXt (right)}
\label{fig10}
\end{figure}

ResNeXt (Fig. 10) \cite{xie2017aggregated} proposed the concept of cardinality to represent the size of transformations. They designed a block consisting of homogeneous, multi-branches and a skip connection. Experimental results on the ImageNet-1K dataset demonstrate that increasing cardinality is more effective and efficient than simply increasing the depth and width. In addition, multi-scale features also play an important role in various vision tasks. To this end, a novel multi-scale backbone architecture named Res2Net was proposed in \cite{gao2019res2net} (Fig. 11). The multi-scale residual features are effectively learned from the  Res2Net module. Unlike the previous blocks in ResNet and ResNeXt, Res2Net incorporates hierarchical residual-like connections within a single residual block to better represent the multi-scale features in one residual block.

\begin{figure}[http]
\centering
\includegraphics[width=3.5in]{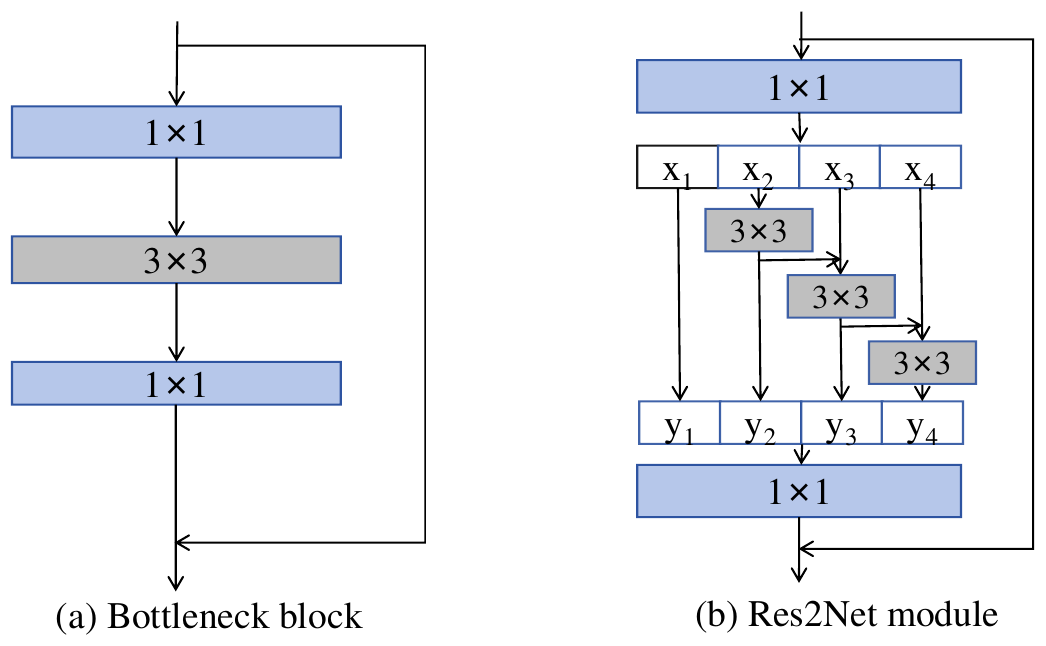}
\caption{A residual learning block (left) and a basic block from Res2Net (right).}
\label{fig11}
\end{figure}

While the aforementioned ResNeXt and Res2Net widen the feature map by adding more branches, the wide residual networks (WRNs, Fig. 12) achieve the same purpose by increasing the number of feature maps in one residual block \cite{zagoruyko2016wide}. It reduces the total depth of ResNet and widens the  feature maps in each residual block, showing its superiority over thin and deep counterparts.

\begin{figure}[http]
\centering
\includegraphics[width=3.5in]{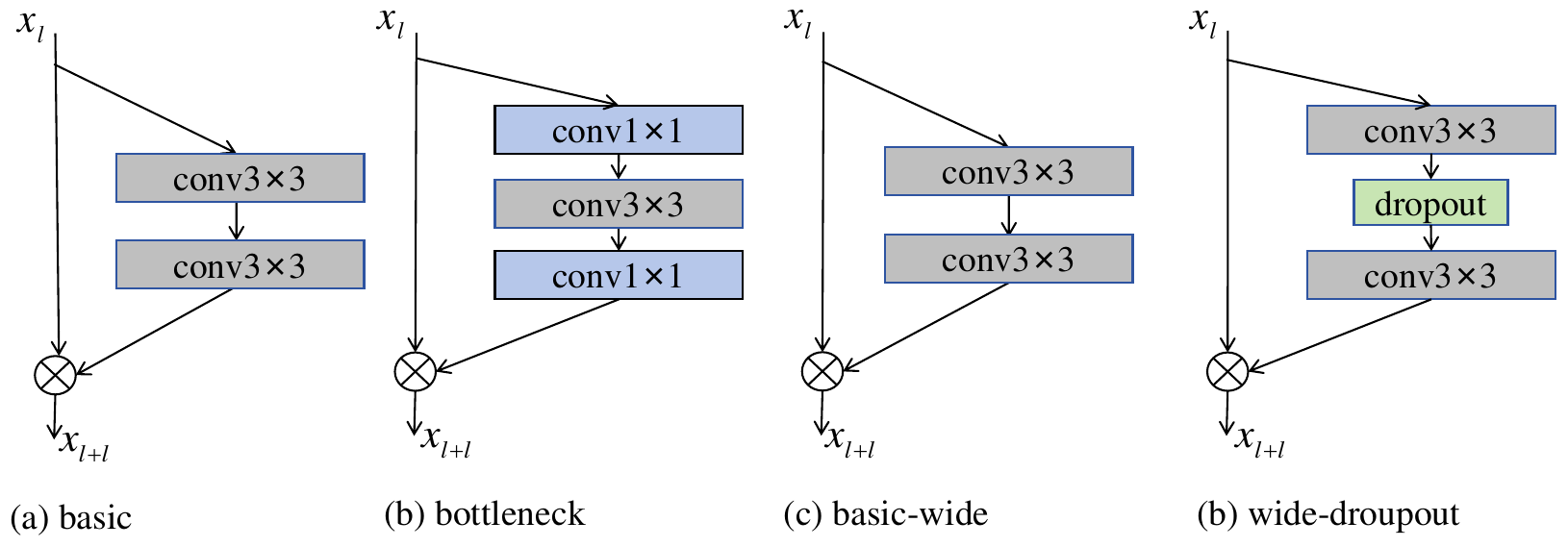}
\caption{Four kinds of residual learning blocks. Here the word “wide” means the output of the first convolution block has a greater number of feature maps compared to the initial residual block.}
\label{fig12}
\end{figure}

In short, widening the residual units could improve the final performance in terms of efficiency and accuracy. Two of the most popular strategies to achieve this are introducing  multiple branches and increasing the number of feature maps in a residual block. 

\begin{figure}[http]
\centering
\includegraphics[width=3.5in]{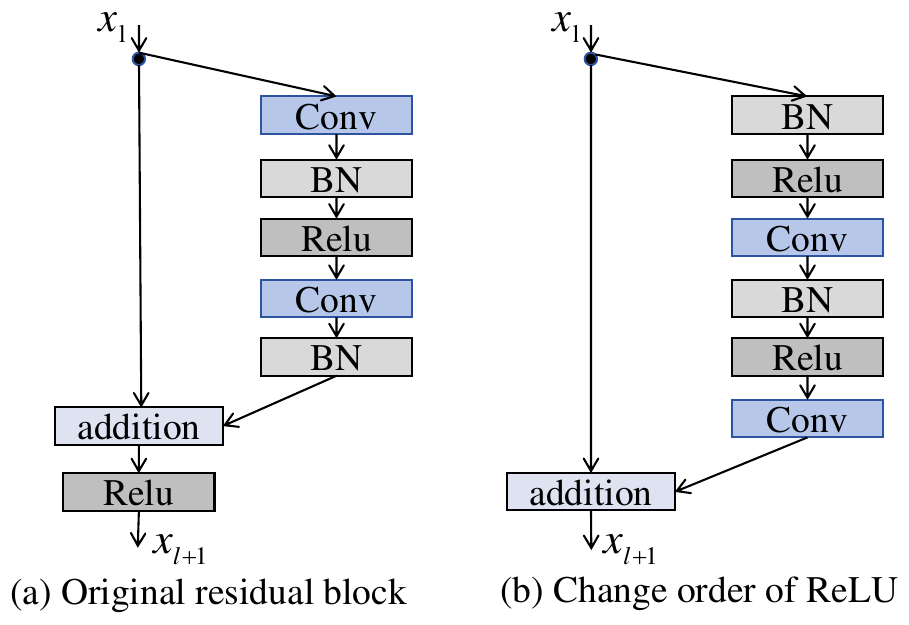}
\caption{Change the order of ReLU in the residual block.}
\label{fig13}
\end{figure}

\subsection{Strengthen the ability to learn discriminative features in residual units}
Residual learning has been found to be effective and simple due to the introduction of skip connections. Many researchers have developed advanced modules to strengthen the ability, aiming to learn discriminative residual features. In this survey, we present several important and popular techniques to this end, along with representative works that utilize these techniques.
\subsubsection*{\bf The order of activation function in the residual block}
In the original residual block, the activation function is applied after the addition of residual features and identity features. However, in \cite{he2016identity}, the authors found that using the pre-activation method in the residual unit allows for direct propagation of forward and backward signals between blocks (see Fig. 13). The proposed new residual block improves training generalization.

\begin{figure}[http]
\centering
\includegraphics[width=3.5in]{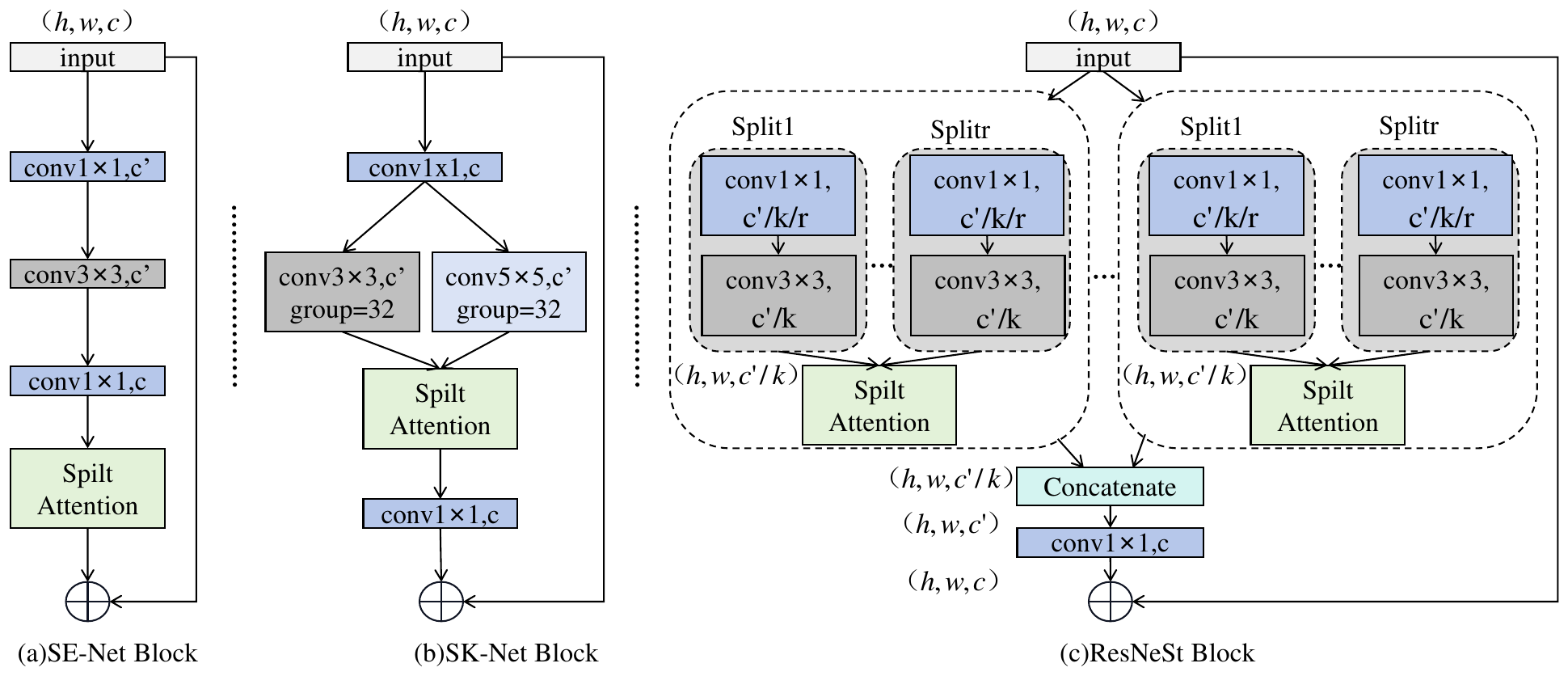}
\caption{ Block of SENet (left), SKNet (middle) and ResNeSt (right)}
\label{fig14}
\end{figure}

\subsubsection*{\bf  Attention mechanism in the residual block}
The attention mechanism is a method for identifying significant regions in the spatial dimension or the importance of feature maps in the channel dimension. It assists neural networks in focusing more on important information by assigning more weight to it. In SENet \cite{hu2018squeeze}, a squeeze-and-excitation block was proposed (Fig. 14, left), which can learn the attention weights for each channel  adaptively by explicitly modeling interdependencies between each input feature map. Afterward, the original features are added to the weighted feature maps via a skip connection. In SKNet \cite{li2019selective}, the input feature maps are divided into two groups and convoluted with different kernel sizes to achieve multi-scale features. Then, channel-attention is employed to learn the importance of each feature map. Finally, the identity maps are added to the residual features with skip connections (Fig. 14, middle). In ResNeSt \cite{zhang2022resnest}, a multi-branch architecture was proposed that consists of a series of Split-Attention modules (Fig. 14, right). The Split-Attention module in ResNeXt inherits the concept of cardinal and divides input feature maps into $k$ groups. This allows for the calculation of channel weights by a split attention block in each cardinal group.  Finally, similar to SKNet and SENet, the input feature maps are added with the channel-wise weighted residual feature maps.

\begin{figure}[http]
\centering
\includegraphics[width=3.5in]{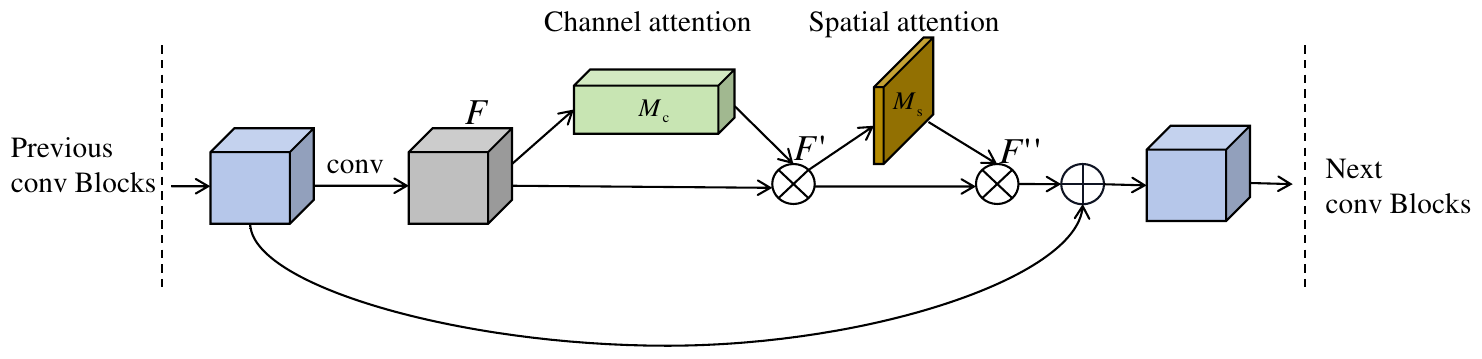}
\caption{ The block of CBAM.}
\label{fig15}
\end{figure}

In addition to channel-wise attention, spatial attention can also be integrated into residual blocks to learn rich representations. Two types of attention modules based on channel and spatial attention are proposed for adaptive feature refinement in \cite{woo2018cbam} and \cite{fu2019dual}. The channel attention and spatial attention modules are connected in series \cite{woo2018cbam}, with the skip connection introduced after the spatial attention (Fig. 15). Unlike the CBAM in \cite{woo2018cbam}, the position (spatial) channel module and channel attention module are connected in a parallel way, and skip connection is used in each module (Fig. 16).

In short, the primary motivation for incorporating an attention mechanism into the residual block is to assist the network in capturing comprehensive contextual information and guiding the learning of discriminative features. 

\begin{figure}[http]
\centering
\includegraphics[width=3.5in]{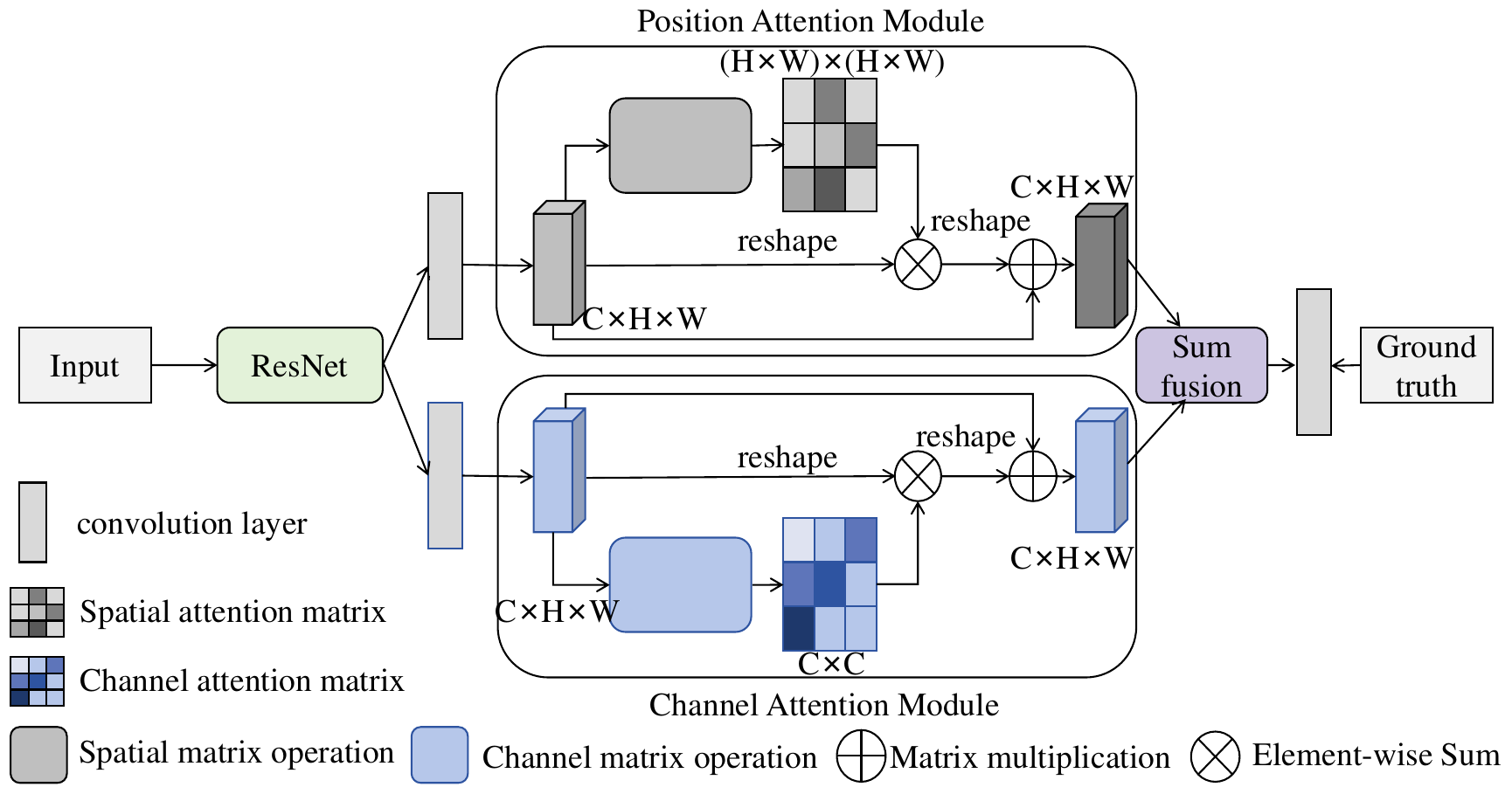}
\caption{ The Architecture of Dual Attention Network}
\label{fig16}
\end{figure}

\subsubsection*{\bf  Rethink bottleneck structure in the residual block}
The original ResNet-50/101/152 utilized a bottleneck residual block, which includes a 1×1 layer for reducing feature map dimension, a 3×3 layer for feature learning, and a 1×1 layer for restoring dimension  (Fig. 17 (a)). In MobileNetv2 \cite{sandler2018mobilenetv2}, an inverted bottleneck residual block is used instead (Fig. 17 (b)). In contrast to the regular bottleneck structure, the inverted bottleneck connects the feature maps with a low number of channels at both ends of the residual path, and lightweight depthwise convolutions are adopted for the intermediate expansion layer. The inverted bottleneck provides a separation between the input/output dimensions and the layer transformation by detangling the capacity and expressiveness of the network. Note that the skip connection is also used in these two types of bottlenecks by adding both ends of feature maps.

\begin{figure}[http]
\centering
\includegraphics[width=3.5in]{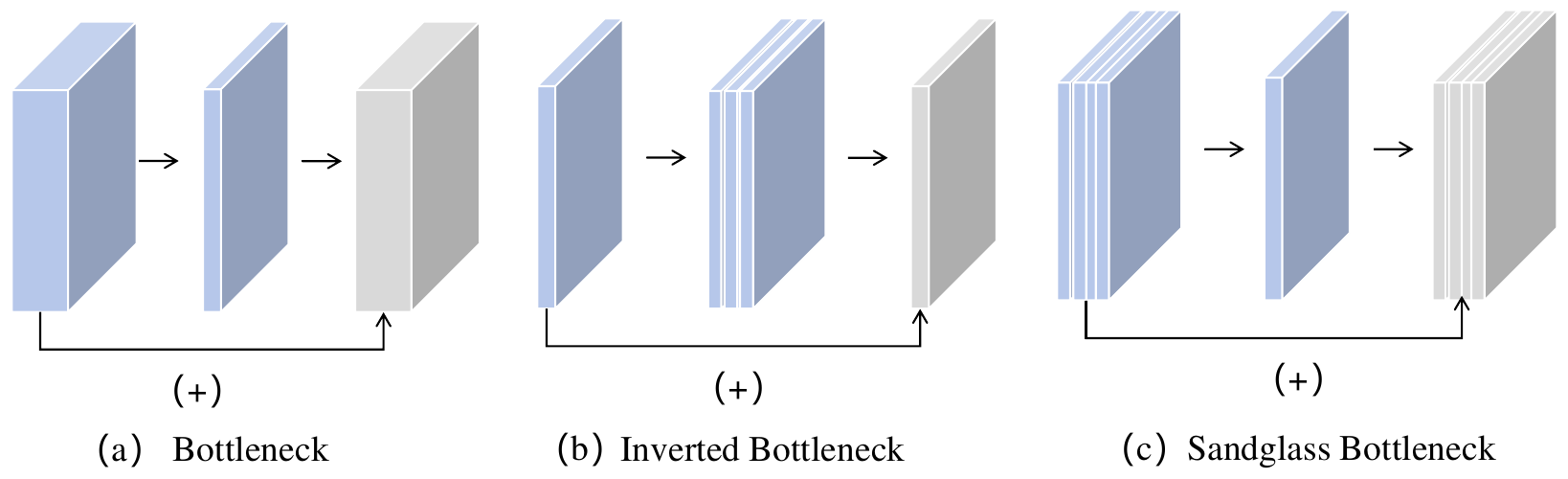}
\caption{ Three types of bottlenecks in the residual block}
\label{fig17}
\end{figure}

In their study \cite{zhou2020rethinking}, the authors discovered that the inverted bottleneck structure may lead to  information loss and gradient confusion. To address this issue, they proposed a sandglass bottleneck block (Fig. 17 (c)), which utilizes depthwise convolutional operations at both ends of the regular residual bottleneck. Their experimental results demonstrate that the proposed bottleneck structure offers more advantages in terms of accuracy and trainable parameters compared to the inverted ones. In addition, a lightweight GhostNet was proposed by stacking ghost bottlenecks \cite{han2020ghostnet}. The ghost bottleneck generates more feature maps from cheap linear transformation operations, based on the belief that the redundancy in feature maps is essential for boosting performance. 
In a nutshell, most works on bottleneck structure focus on how to balance performance and computational resources for efficient deep-learning architectures. 

\begin{figure}[http]
\centering
\includegraphics[width=2.5in]{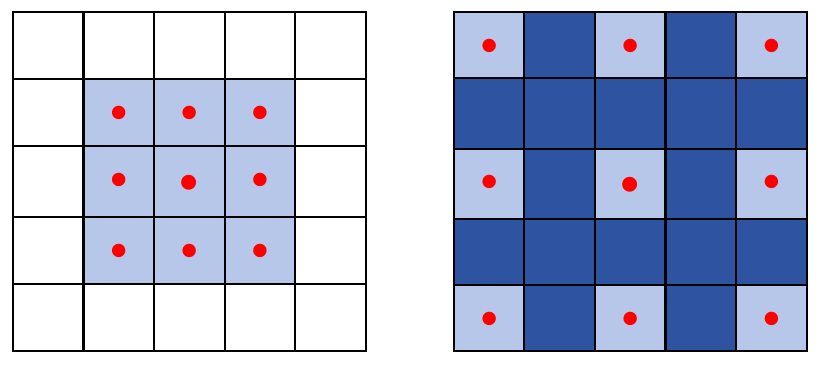}
\caption{ Illustration of dilated convolution. Left: standard convolution with kernel size 3 × 3; Right: dilated convolution with a kernel size of 3 × 3, a dilation rate of r = 2, which enlarges the reception field to 5 × 5. The dark blue boxes represent zeros that are inserted during dilated convolution. Please note that the color references in this figure legend can be found in the web version of this article.}
\label{fig18}
\end{figure}

\subsubsection*{\bf  Dilated and deformable convolution in the residual block}
Thanks to the advantages of dilated and deformable convolutions for learning multi-scale features and contextual information, many researchers are exploring their integration  into residual blocks.
Dilated (atrous) convolution was proposed and developed in the DeepLab series \cite{chen2014semantic}\cite{chen2017deeplab}\cite{chen2017rethinking}\cite{chen2018encoder}  for semantic segmentation. It enlarges the kernel by inserting holes (zeros) between the kernel elements. An additional parameter, the dilation rate, indicates how much the kernel is widened (Fig. 18). By using different dilation rates, we can achieve multi-scale features without sacrificing the resolution of feature maps, which helps to improve the segmentation performance. 
Dilated convolutions embedded in residual blocks are widely used in various tasks, such as image classification \cite{yu2017dilated}, depth estimation \cite{chabra2019stereodrnet}, MRI reconstruction \cite{dai2019compressed}, image denoising \cite{jia2023multi}, and image segmentation \cite{raju2021multi}. In the ESPNet \cite{mehta2018espnet} and ESPNetv2 \cite{mehta2019espnetv2}, an efficient spatial pyramid of dilated convolutions integrated into skip connections is proposed for semantic segmentation(Fig. 19). In \cite{yu2017dilated}, a dilated residual network (DRNs) was proposed and shown to outperform the non-dilated counterparts in image classification, object localization and semantic segmentation without increasing the model’s complexity. 

\begin{figure}[http]
\centering
\includegraphics[width=3.5in]{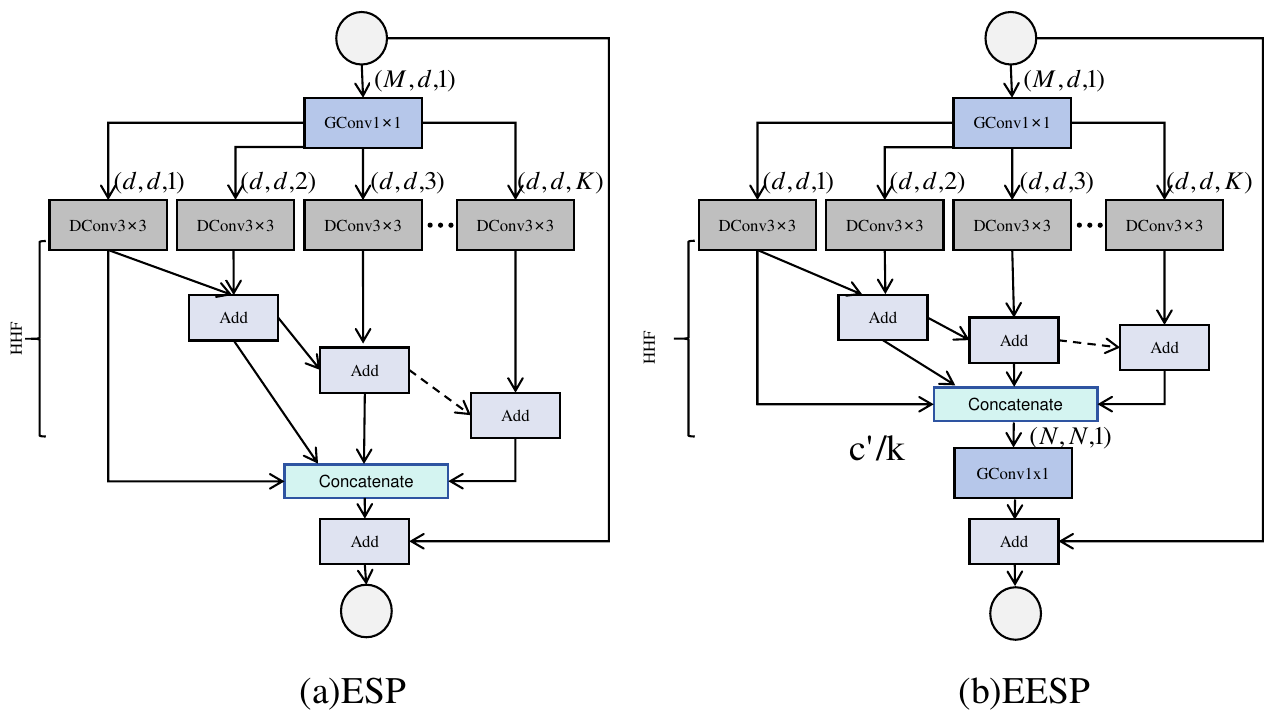}
\caption{  The block of ESP and EESP for ESPNet and ESPNetv2, in which the dilated convolution is used to learn residual features}
\label{fig19}
\end{figure}

\begin{figure}[http]
\centering
\includegraphics[width=3.5in]{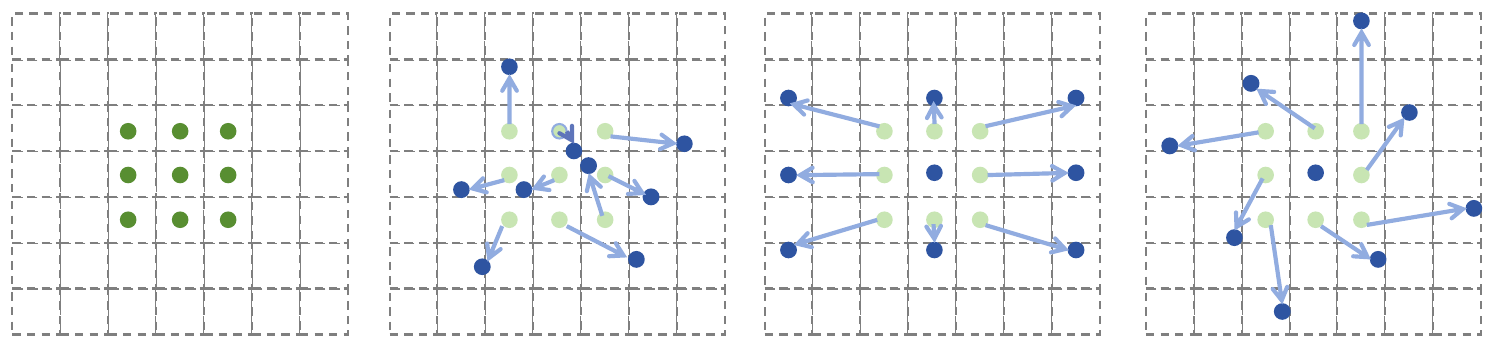}
\caption{ The regular convolution, deformation convolution and dilated convolution}
\label{fig20}
\end{figure}

Deformable convolution networks \cite{dai2017deformable} aim to enhance the transformation modeling capability of regular convolution operations. They achieve this by adding an additional convolutional layer to learn 2D offsets for the regular grid sampling locations in the standard convolution. Thus, this allows the convolution operation to focus not only on a local window but also in an adaptive way based on the input features (Fig. 20.). Several researchers have explored the possibility of employing deformable convolution in residual structures. Their results have shown that adding skip connections in the deformable convolution module can enhance the feature learning capability in various applications, including video action segmentation \cite{lei2018temporal}, image super-resolution \cite{zhang2022deformable}\cite{ying2020deformable}, hyperspectral image classification \cite{zhang2021spectral}, and retinal vessel segmentation \cite{yang2022dcu}.

In summary, the dilated convolution and deformable convolution are operations that enhance the capabilities of the standard convolution, which could help achieve multi-scale feature maps and focus on semantic objects, respectively. When integrated into a residual unit, they can improve discriminative feature learning in deep neural networks.

\begin{figure}[http]
\centering
\includegraphics[width=3.5in]{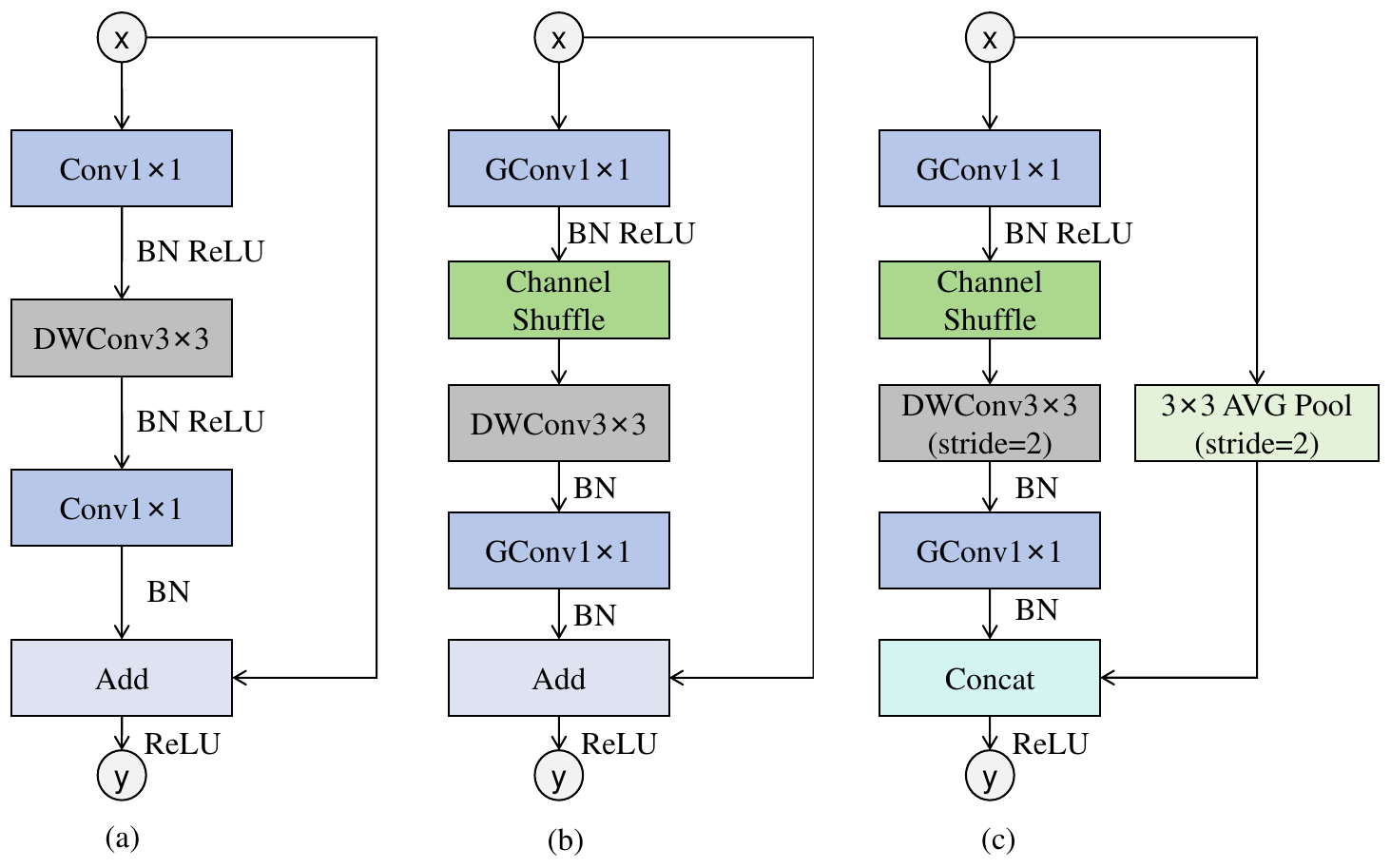}
\caption{ ShuffleNet blocks. a) bottleneck block with depthwise convolution (DWConv); b) ShuffleNet unit with pointwise group convolution (GConv) and channel shuffle; c) ShuffleNet block with stride 2 for downsampling the spatial resolution of feature maps.}
\label{fig21}
\end{figure}

\subsection{Make the ResNet-like more efficient}
Although residual learning with skip connections allows for training deep convolutional networks without performance degradation, a major issue is the relatively slow training time as the networks become deeper. Efficient deep learning has recently become a popular topic due to the interest in edge devices, such as smartphones, robots, autopilots, etc. Efficient deep learning could be achieved by balancing accuracy and efficiency in ResNet-like models. Since convolution layers in DCNNs contribute to most of the computation, designing efficient convolution operations is crucial for building efficient CNNs. The three widely used convolution operations are 1x1 convolution, group convolution, and depthwise convolution(Fig. 21) \cite{cai2022enable}. This section will focus on these three techniques that, when combined with skip connections, facilitate efficient learning.

\begin{enumerate}

\item{1x1 convolution, also called pointwise convolution, uses a kernel size of 1. This operation reduces multiply-accumulate operations (MACs) and training parameters by K2 times compared to a standard KxK convolution. In NIN \cite{lin2013network}, the 1x1 convolution is utilized after standard convolution to improve the model’s ability to distinguish local patches within the receptive field. Inspired by NIN, GoogLeNet \cite{szegedy2015going} (also called Inception-v1) introduced 1x1 convolutions to reduce dimensionality and increase the width of feature maps in a single layer. Similarly, in ResNet, 1x1 convolutions are used  in bottleneck design for accelerating training speed. Specifically, for a residual block, three convolutional layers are used instead of one, consisting of 1x1, 3x3, and 1x1 convolution. }
\item{Group convolution involves performing the convolution operation on each group of feature maps. In AlexNet, the feature maps were divided into two groups, and the convolution operation was performed on each group separately on a single GPU due to the limited GPU memory. In ResNeXt \cite{xie2017aggregated}, the group convolution is integrated into the residual block, demonstrating the effectiveness of this structure. In ShuffleNet \cite{zhang2018shufflenet}, pointwise group convolution and channel shuffle were introduced to the residual block for reducing computation overhead while maintaining accuracy. }
\item{Depthwise convolution is an extreme case of group convolution, where the groups $G$ are equal to the number of input channels. Due to the lack of interactivity between feature maps, the modeling capacity is lower than the group or normal convolution. In practice, the depthwise convolution is usually combined with 1x1 convolution. The former is used to learn spatially in each feature map, while the latter combines the outputs from the depthwise convolution. In MobileNet \cite{howard2017mobilenets} and Xception \cite{chollet2017xception}, the combination of depthwise convolution and pointwise convolution (also known as depthwise separable convolution) becomes the basic module for efficient deep learning. In MobileNetV2 \cite{sandler2018mobilenetv2} and MobileNetV3 \cite{howard2019searching}, the depthwise separable convolution is employed in the residual block (Fig. 22). }    
\end{enumerate}

\begin{figure}[http]
\centering
\includegraphics[width=3.5in]{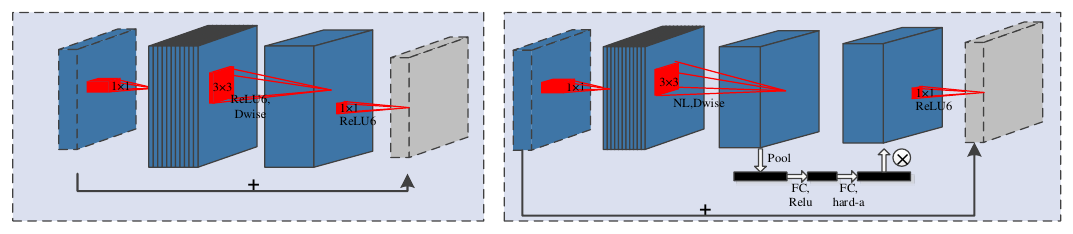}
\caption{ The depthwise convolution in MobileNetV2 (left) and MobileNetV3 (right)}
\label{fig22}
\end{figure}

In addition to the above three efficient convolution methods, some works focus on the redundancy of the residual feature maps. In \cite{huang2016deep}, the authors proposed a stochastic depth training strategy by randomly dropping a subset of residual units and bypassing them with skip connections (Fig. 23).  In \cite{ResKD}, the residual-guided knowledge distillation was proposed to learn a lightweight and efficient student network. A way of pruning residual connections was proposed via a KL-divergence-based criterion in \cite{Luo_2020_CVPR}.
\begin{figure}[http]
\centering
\includegraphics[width=3.5in]{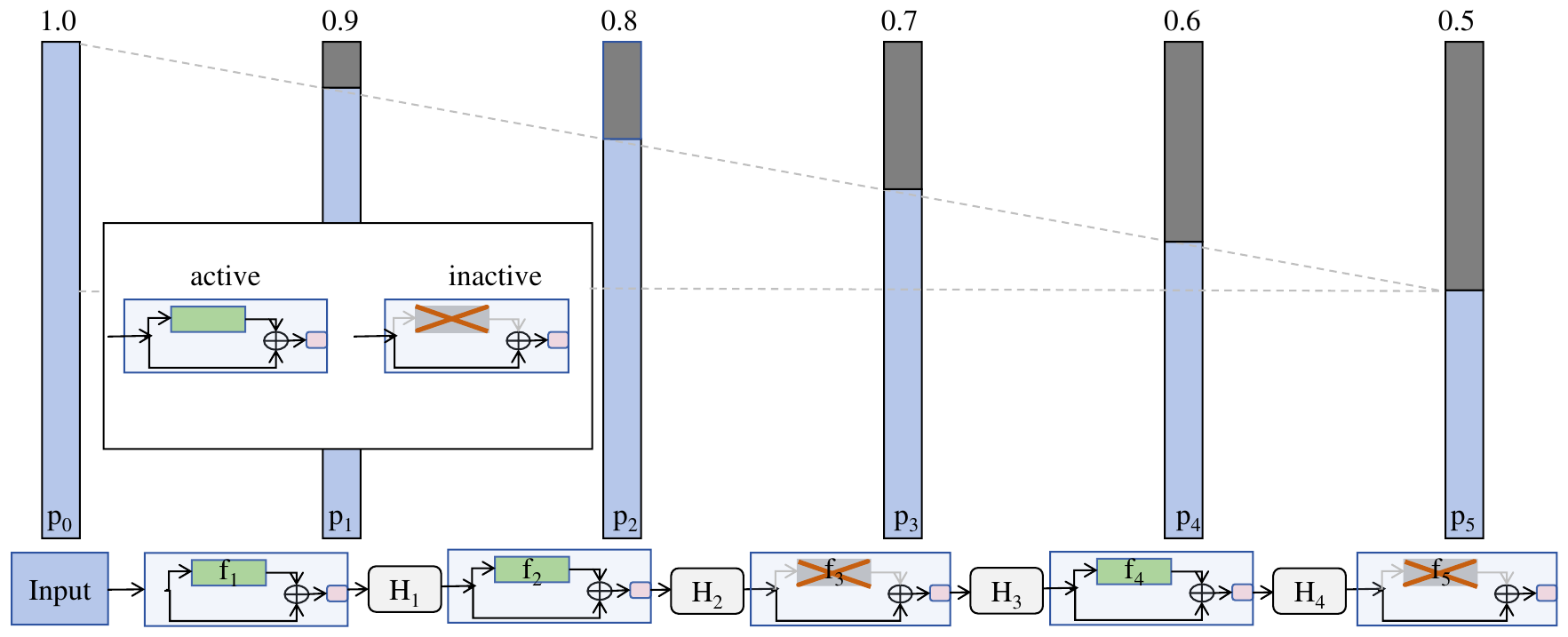}
\caption{ the structure of stochastic depth with randomly dropping a subset of residual units}
\label{fig23}
\end{figure}

\subsection{Skip connection in self-attention}
Convolutional Neural Networks (CNNs) encounter difficulties capturing the contextual information of entire images due to their locally biased receptive fields. To overcome the limitations imposed by fixed receptive fields in CNNs, Wang et al. \cite{wang2018non} incorporated self-attention into CNNs with a proposed non-local block (Fig. 24). The proposed spacetime non-local block utilizes 1x1x1 convolutions to transform input features, capturing long-range dependencies through softmax. After the operation g, the reweighted features underwent a subsequent 1x1x1 convolution to restore channel numbers. Importantly, the initial input features were then reintroduced via skip connections.

\begin{figure}[http]
\centering
\includegraphics[width=3.5in]{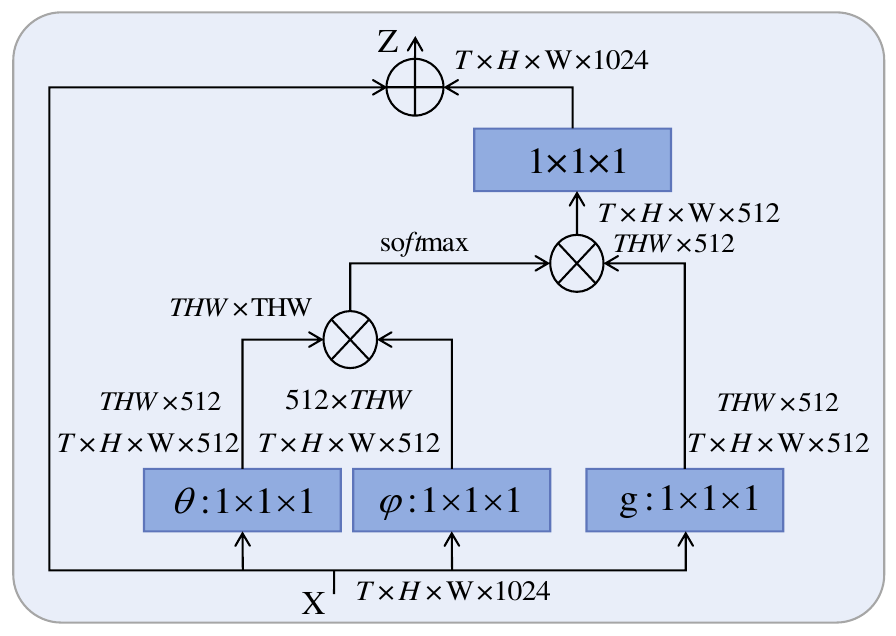}
\caption{ A spacetime non-local block}
\label{fig24}
\end{figure}

While self-attention has shown excellent performance in building long-range dependencies in the whole feature maps, its quadratic complexity of the self-attention for each position limits its application in computer vision. Some works have attempted to overcome this limitation and accelerate the speed of self-attention calculation. In CCNet \cite{huang2019ccnet}, the authors treated self-attention as a graph convolution, introducing criss-cross attention to reducing complexity from \(O(n^2)\) to \(O(n\sqrt{n})\). The EMANet \cite{li2019expectation} employs the expectation maximization (EM) algorithm to obtain compact bases, reducing complexity from \(O(n^2)\) to \(O(nk)\), where \(k\) represents the number of compact bases. Both studies also adopt skip connections, which add input features to attention residual features.

Due to the computational complexity directly related to the number of pixels in feature maps, self-attention is typically introduced in the last few layers of CNNs. However, this limits the model's ability to extract fine-grained features, which are essential for specific vision tasks, such as semantic segmentation, object detection, and super-resolution reconstruction. In 2017, a novel framework based solely on self-attention mechanisms, named Transformer \cite{vaswani2017attention} (Fig. \ref{fig26}), was proposed for machine translation tasks. The skip connections in the basic module of the Transformer are also utilized for residual feature learning.

\begin{figure}[http]
\centering
\includegraphics[width=2.0in]{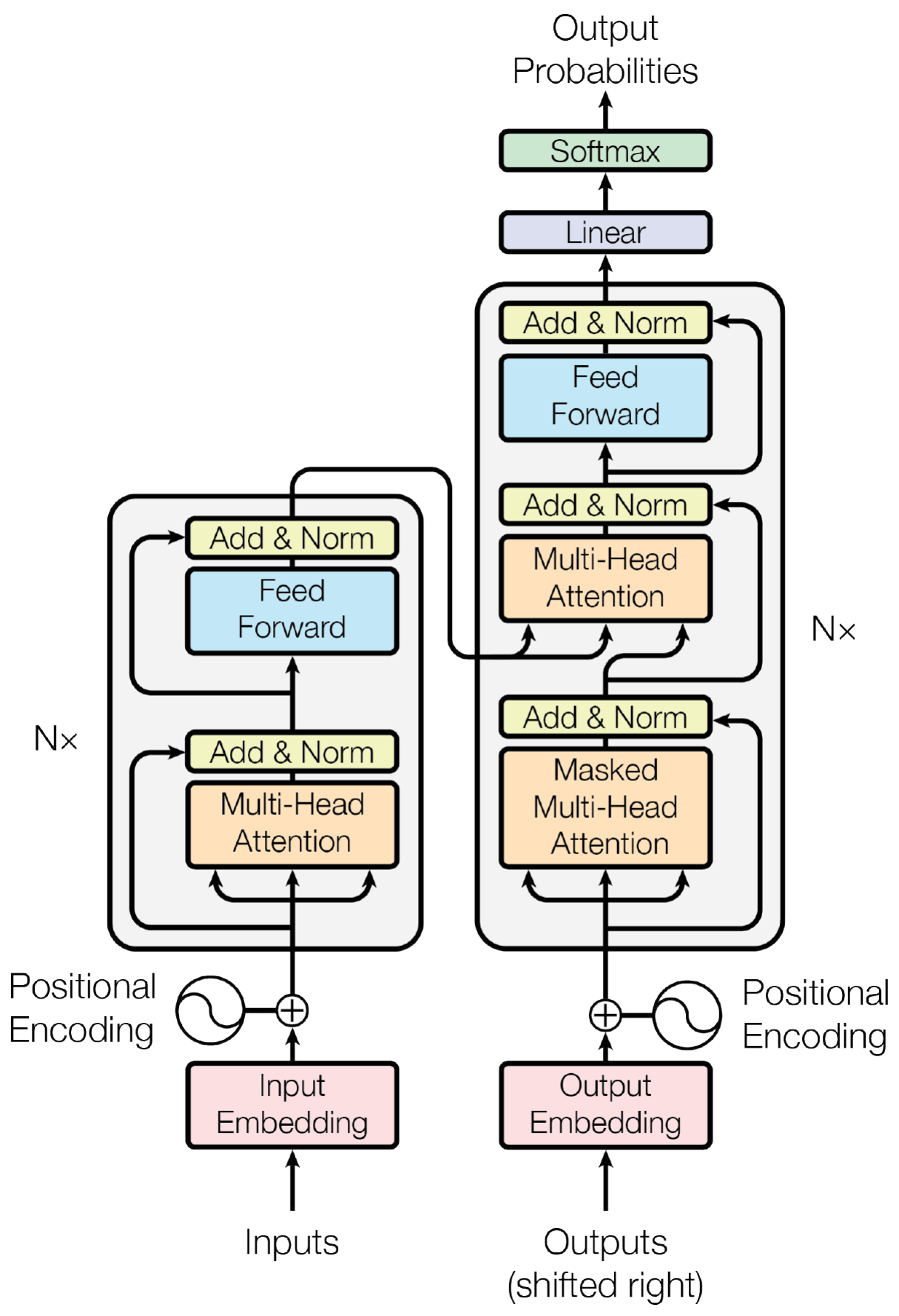}
\caption{The architecture of Transformer. From \cite{vaswani2017attention}}
\label{fig25}
\end{figure}

With the remarkable success of the Transformer in natural language processing (NLP), some researchers have attempted to apply it to computer vision tasks. The Vision Transformer (ViT) was introduced in \cite{dosovitskiy2020image}. It simulated the way of NLP approach of dividing an image into fixed-size patches, linearly embedding each patch, adding position embeddings, and then feeding the resulting sequences into a Transformer encoder (Fig. 26). It has been demonstrated that a pure transformer can be directly applied to a sequence of image patches with the help of pretraining on a large amount of image data. 

\begin{figure}[http]
\centering
\includegraphics[width=3.5in]{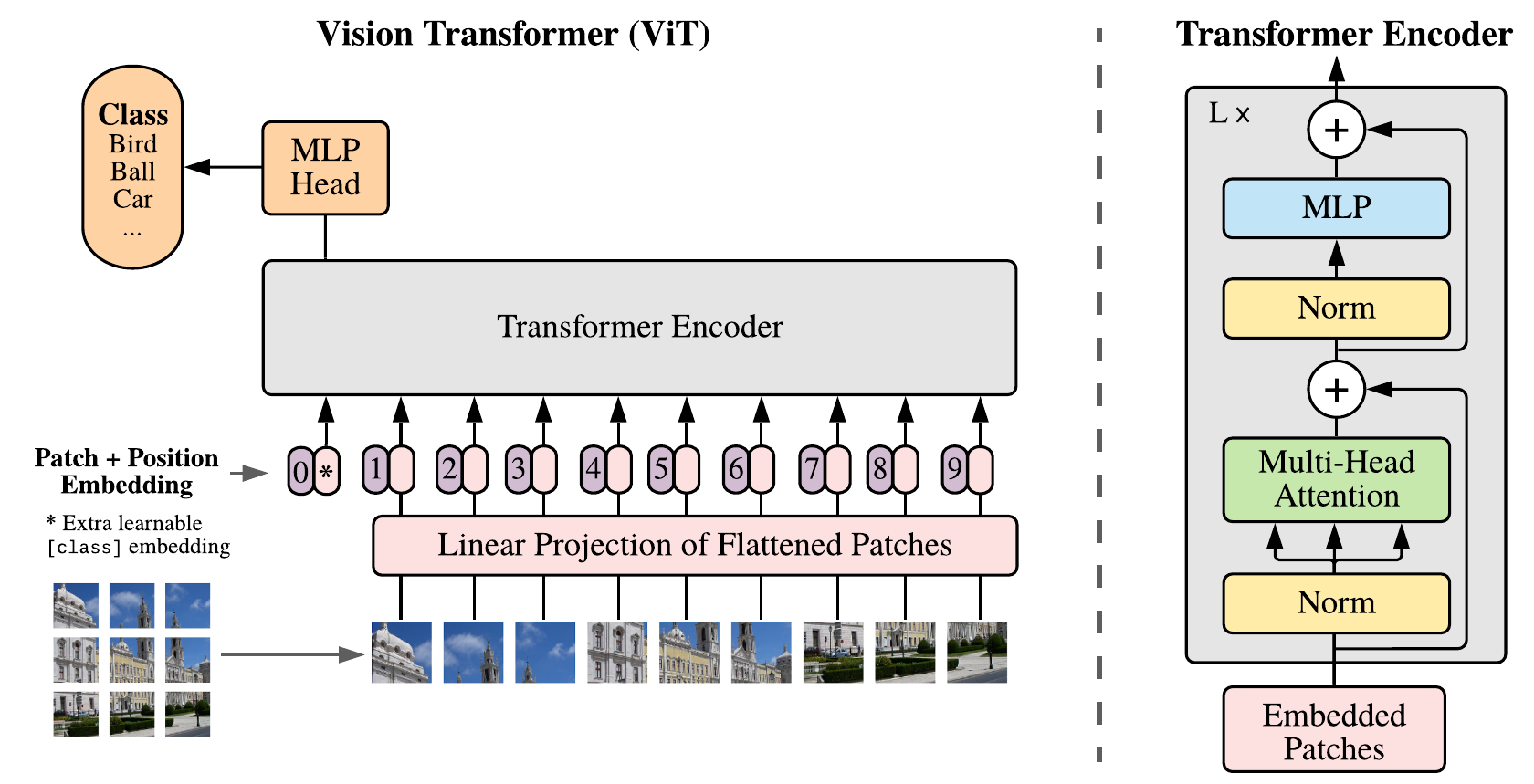}
\caption{The architecture of the Vision Transformer (left) and the Transformer Encoder (right) From \cite{dosovitskiy2020image}}
\label{fig26}
\end{figure}

Considering the difference between language and vision, such as the large-scale variations of visual entities compared to words, a hierarchical pure self-attention called Swin-Transformer was proposed in \cite{liu2021swin}. The main structure of the Swin Transformer block is shown in Fig. 27. The model uses a windowing scheme to compute self-attention in local windows, reducing computing overhead.,. Meanwhile, the long-range dependency relationship can also be learned effectively using the hierarchical and shifted window operations. The skip connection is also utilized in self-attention (W-MSA and SW-MSA) and multiple-layer perceptions (MLP).

\begin{figure}[http]
\centering
\includegraphics[width=2.5in]{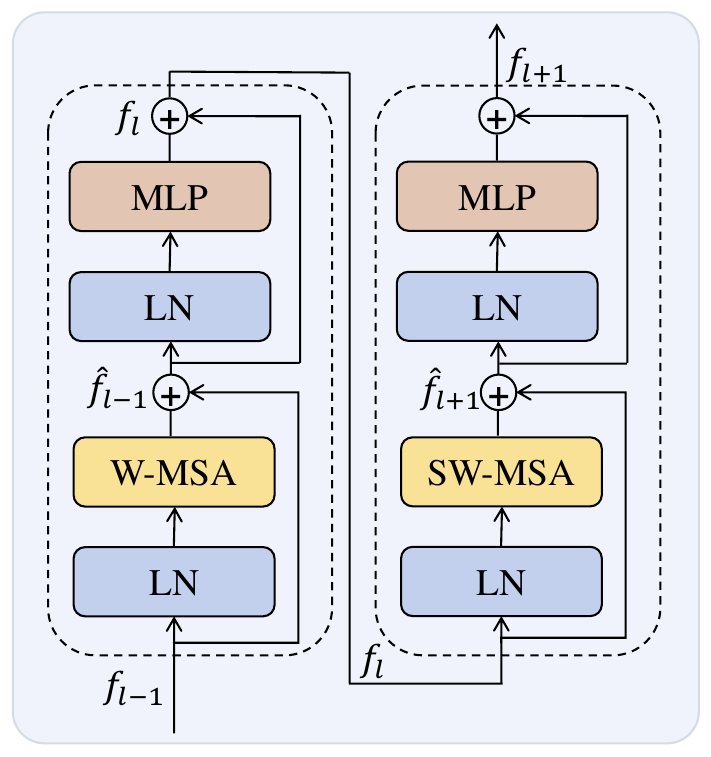}
\caption{The structure of the Swin Transformer block. Here, W-MSA and SW-MSA refer to window multi-head shifted attention and shifted window multi-head shifted attention, respectively}
\label{fig27}
\end{figure}

The success of Transformer in computer vision has led to residual learning being more popular. The skip connection has become  the default choice for designing various Transformer architectures, including MoCo v3\cite{mocov3} and MAE\cite{mae}, for self-supervised pretraining, DETR \cite{carion2020end}, and Deformable DETR \cite{zhu2020deformable} for object detection, SETR \cite{zheng2021rethinking}, and TransUNet \cite{chen2021transunet} for segmentation, and TransGAN \cite{jiang2021transgan} and DALLE-E \cite{ramesh2021zero} for image generation etc. 

\section{Explanation of skip connection in the residual block }
\label{sec5}
Due to the success of skip connections in residual blocks, several works have attempted to dive into the theoretical background behind residual learning. In this section, we survey some explanations for why residual learning is effective in deep neural networks.
\subsection{Information flow }
Numerous studies have creatively utilized residual connections in diverse ways, enhancing information flow processing and optimization. Wang et al. \cite{9754543} accomplished fine estimation of scene flow by explicitly learning residual scene flow instead of embedding features, which had a direct compensatory effect on long-distance motion estimation. Hosny et al. \cite{hosny2022refined} proposed a model that utilizes residual learning techniques to address image scarcity and degradation issues by improving information flow and reconstructing layers. Meanwhile, Chang et al. \cite{chang2022strpm} efficiently extract spatiotemporal residual features through residual connections, with a focus on residual prediction memory for frame-to-frame motion modeling in high-resolution videos. Anwar et al. \cite{anwar2019real} use residuals to mitigate low-frequency information flow and apply feature attention mechanisms to exploit channel dependencies. Zhang et al. \cite{zhang2018residual} maximize the utilization of layer information in each block through dense connections while further optimizing information flow and gradients through local residual learning, providing a more comprehensive consideration for the model's information processing.

\subsection{Ensemble learning}
In computer vision, the conventional processing sequence typically involves gradually extracting low-level features into high-level features. However, the residual method in ensemble learning has gained widespread attention and development. The introduced identity skip connections bypass residual layers, allowing any layer to flow directly to any subsequent layer. In the context of ensemble learning, residual learning improves the overall system's robustness to noise and outliers by capturing residual information between models. For instance, Zhang et al.  \cite{veit2016residual} introduced a progressive ensemble framework that effectively enhances the stability and noise resistance of the learning system by using the residuals of the previous base classifier to train the next one. Their work demonstrates that residual networks can be viewed as a collection of multiple paths that create implicit connections between the input and output, as illustrated in Fig.28. Disrupting these paths has only a moderate impact on performance, highlighting the essential role of residual learning in improving system robustness and resistance to interference. The variety of these pathways reinforces the model's robustness, allowing it to more effectively adjust to intricate environments and evolving data.

\begin{figure}[http]
\centering
\includegraphics[width=3.5in]{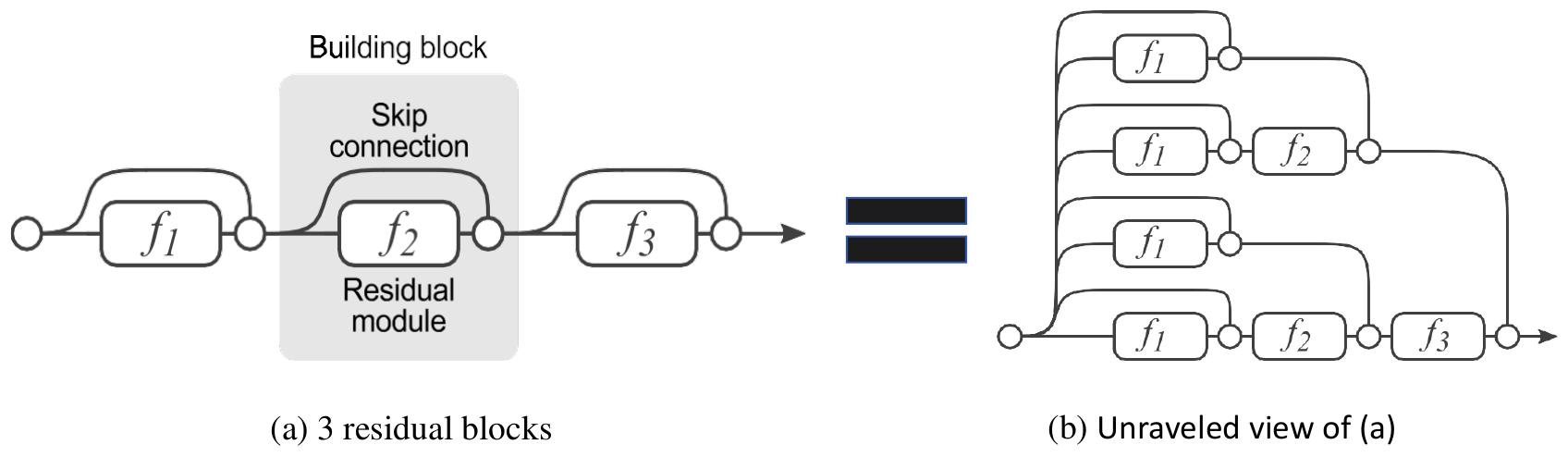}
\caption{Ensemble learning: The basic block of ResNets is shown as (a), which can be regarded as an unraveled view of  (b). It can be seen as adding a residual block doubles the number of paths. From \cite{veit2016residual}}
\label{fig28}
\end{figure}

\subsection{Regularizations}
Regularization aims to prevent models from overfitting to training data, thereby improving their generalization ability. Residual blocks have been shown to facilitate the optimization process and enhance the representational capacity of networks. Residual learning plays a crucial role in the field of regularization, as it effectively controls the complexity of models, enabling them to better adapt to unseen data. The Shake-shake operation \cite{gastaldi2017shake} in a three-branch residual network involves adjusting the multiplication factors of each branch during forward propagation and multiplying them by different values during backward propagation, thereby altering the influence of each branch on the final result. This residual learning mechanism allows the model to adjust parameters and better adapt to data distributions flexibly. ShakeDrop \cite{yamada2019shakedrop} is an improvement on Shake-shake that applies residual learning. Regarding regularization, residual learning introduces shortcut paths through skip connections, which helps the model learn the differences between input and output more easily. Additionally, according to \cite{gao2019vacl}, placing the filters of skip connections into regularization groups and introducing variance-aware cross-layer regularization to constrain intra-group variance effectively enhances the structural sparsity of residual models. In \cite{li2018visualizing}, the authors proposed a filter normalization scheme to visualize the loss surface. They found the skip connections could smooth the loss surface (See Fig. 29), suggesting that the skip connections could play as regularizations in training a deep neural network. 

\begin{figure}[http]
\centering
\includegraphics[width=3.5in]{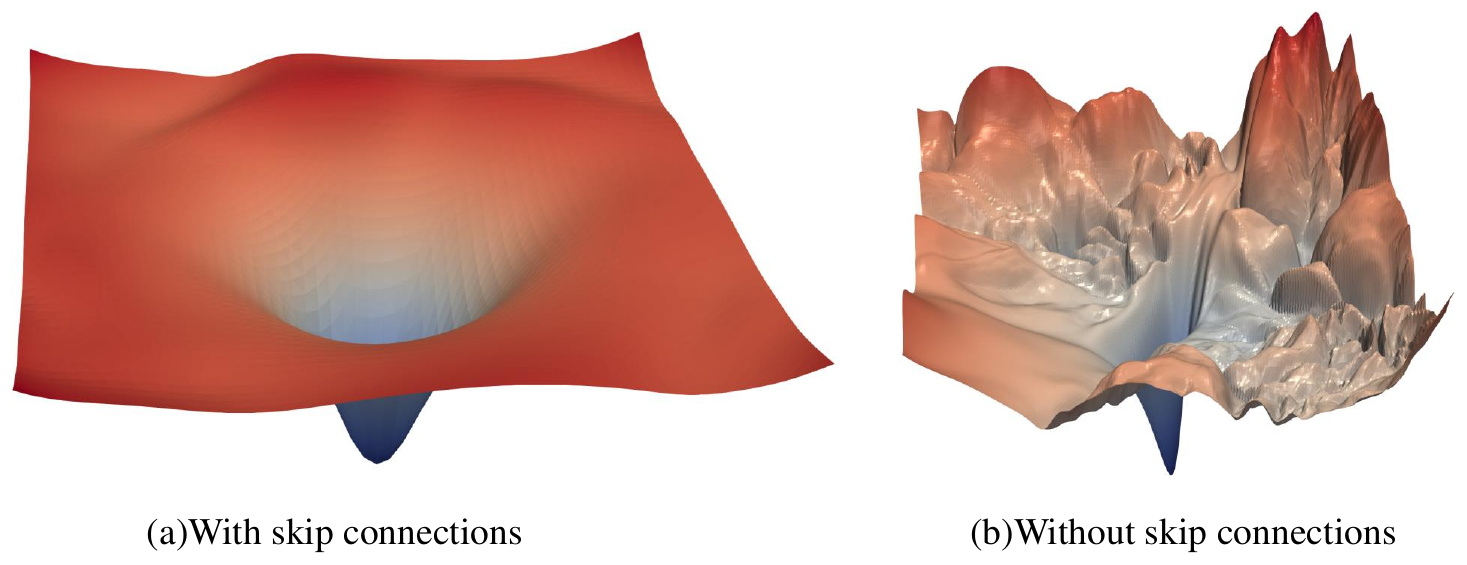}
\caption{The loss surface of ResNet-56 (a) with and (b) without skip connections. From [134]. From \cite{li2018visualizing}}
\label{fig29}
\end{figure}

\subsection{Eliminate singularities}
In the field of deep neural networks, skip connections optimize training to some extent by addressing singularity issues. \cite{orhan2017skip} confirms that skip connections improve training outcomes, particularly alleviating the learning slowdown caused by singularities. The design of skip connections ensures that adjacent layer units remain active for certain inputs, even when their adjustable connections become zero, effectively eliminating singularities (see Fig. 30a). This design breaks the symmetry of hidden unit arrangements, eliminates overlapping singularities (see Fig. 30b), and prevents units with the same incoming weights from collapsing onto each other. This is achieved by employing different skip connections that continue to remove ambiguities. Furthermore, skip connections effectively eliminate linearly dependent singularities by adding linearly independent inputs, which are mostly orthogonal, as observed in Fig. 30c. Additional research \cite{balduzzi2017shattered} suggests that residual networks effectively address singularity problems, reduce gradient shattering, and enhance numerical stability and optimization effectiveness. The incorporation of skip connections in neural network architectures \cite{yasrab2019srnet} strengthens information propagation, reduces singularities, and further improves performance. Additionally, \cite{oyedotun2022everyone} has shown that deep neural networks with skip connections and residual networks with aggregated features can avoid singularity issues when increasing model depth. This preserves comprehensive information that is beneficial for improved optimization and generalization.

\begin{figure}[http]
\centering
\includegraphics[width=3.5in]{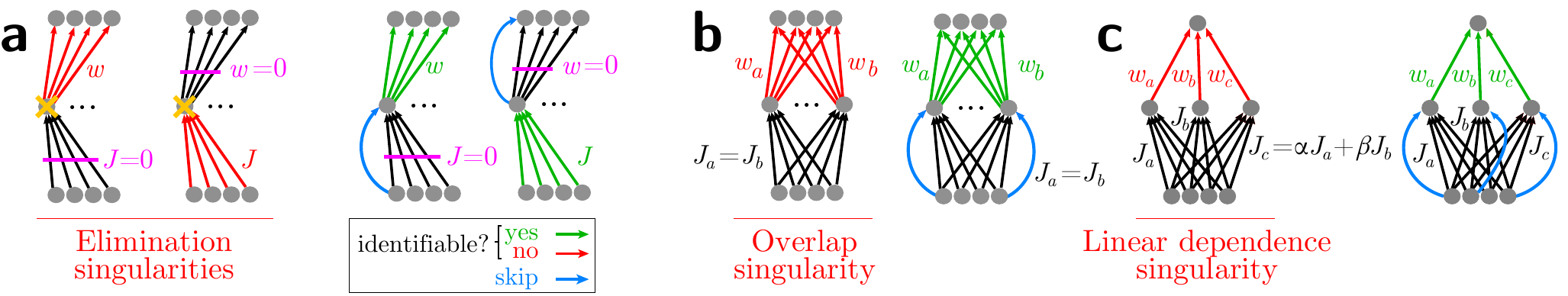}
\caption{The illustration of how skip connections help eliminate singularities. From \cite{orhan2017skip}}
\label{fig30}
\end{figure}

\subsection{Keep gradient stable}
Deep learning has faced a significant obstacle known as the problem of vanishing and exploding gradients. This challenge has been mitigated to a large extent through careful initialization and techniques like batch normalization. However, architectures incorporating skip connections, such as highway networks and residual networks, significantly outperform traditional feedforward structures. Even with careful initialization and batch normalization, this superiority persists. In standard feedforward networks, as depth increases, the correlation between gradients exponentially decays \cite{balduzzi2017shattered} reveals that networks with skip connections exhibit more excellent resistance to gradient shattering and feature sublinear decay.

\section{ skip connection-based methods in computer vision }
\label{sec6}
In this section, we summarized some essential deep neural network architectures that utilize residual learning or use ResNet-like as a backbone for the task of image classification, object detection, segmentation adn image reconstruction in table 1, 2, 3 and 4, respectively. Each table includes the reference, method, link of source code, dataset or image modality, and brief descriptions of each method.  We hope this summarizing could shed light on future research.

\begin{table*}[htbp]
    \centering
    \label{table1}
    \caption{Deep neural networks based on residual learning for image classification}
    \renewcommand{\arraystretch}{1.5} 
    \label{tab:references}
    \begin{tabular}{|l|l|l|p{5cm}|}
        \hline
        \textbf{Reference} & \textbf{Dataset} & \textbf{Code} & \textbf{Description} \\
        \hline
        
        ResNet \cite{he2016deep} & CIFAR10/100, ImageNet & \href{https://github.com/KaimingHe/deep-residual-networks}{Caffe}, 
        \href{https://github.com/gcr/torch-residual-networks}{Torch}, \href{https://github.com/ppwwyyxx/tensorpack/tree/master/examples/ResNet}{TensorFlow}, 
        \href{https://github.com/raghakot/keras-resnet}{Keras} & The initial work that uses skip connection in deep neural networks \\

        
        Highway Net \cite{srivastava2015highway} & MNIST, CIFAR-100 & \href{https://github.com/c0nn3r/pytorch_highway_networks}{PyTorch}, 
        \href{https://github.com/flukeskywalker/highway-networks}{Caffe}
         & Learn the weighted residual features with gate-mechanism \\

        PyramidNet \cite{han2017deep} & CIFAR10/100 & \href{https://github.com/dyhan0920/PyramidNet-PyTorch}{PyTorch}, 
        \href{https://github.com/jhkim89/PyramidNet-caffe}{Caffe} & Gradually increase the channel number of feature maps at all residual units\\
        
        Wu et. al \cite{wu2019wider}  & ILSVRC 2012 & \href{https://github.com/itijyou/ademxapp}{MXNet} & Discuss the influence of width and depth for ResNets \\

        ResNeXt \cite{xie2017aggregated} & ImageNet-1K/5k, CIFAR10/100 & 
        \href{https://github.com/prlz77/ResNeXt.pytorch}{PyTorch}, 
        \href{https://github.com/apache/mxnet/tree/master}{MXNet}
        & Divide residual block into more branches \\
        
        DenseNet \cite{huang2017densely} & CIFAR10/100, ImageNet & 
        \href{https://github.com/bamos/densenet.pytorch}{PyTorch}, \href{https://github.com/liuzhuang13/DenseNetCaffe}{Caffe},
        \href{https://github.com/miraclewkf/DenseNet}{MXNet},
        \href{https://github.com/LaurentMazare/deep-models/tree/master/densenet}{Tensorflow}
        & Reuse all the previous feature maps with long skip connections \\
        
        Inception-v4 \cite{szegedy2017inception} & ILSVRC 2012 & \href{https://github.com/Cadene/pretrained-models.pytorch}{PyTorch} & Introduce skip connection into Inception module \\
        
        DPN \cite{chen2017dual} & ImageNet, Place365-Standard & \href{https://github.com/huggingface/pytorch-image-models}{PyTorch} & Propose Dual Path Networks to share common features with skip connections \\

        WRN \cite{zagoruyko2016wide} & CIFAR-10, CIFAR-100, SVHN, ImageNet & \href{https://github.com/szagoruyko/wide-residual-networks}{PyTorch} & Increase the number of feature maps in residual bottleneck block \\

        i-ResNet \cite{behrmann2019invertible} & MNIST, CIFAR-10 & \href{https://github.com/jhjacobsen/invertible-resnet}{PyTorch} & Enforce the residual units invertible \\

        Mark et. al \cite{mcdonnell2018training}  & CIFAR-10/100, SVHN & \href{https://github.com/McDonnell-Lab/1-bit-per-weight}{TensorFlow} & Train wide residual networks by using a single bit for each weight \\
        
       Wang et.al \cite{wang2017residual} & CIFAR10/100, ImageNet & \href{https://github.com/tengshaofeng/ResidualAttentionNetwork-pytorch}{PyTorch} & Introduce attention mechanism into residual block \\

        MaxViT \cite{tu2022maxvit} & ImageNet-1K/21k, JFT-300M & \href{https://github.com/google-research/maxvit}{TensorFlow} & Propose a new attention module to learn blocked local and dilated global attention \\

        MLP-Mixer \cite{tolstikhin2021mlp} & ImageNet-1K/21k & \href{https://github.com/google-research/vision_transformer}{JAX/Flax}  & Replace convolution layer with MLP in residual block \\

        ConvNeXt \cite{liu2022convnet} & ImageNet-1K/22k & \href{https://github.com/facebookresearch/ConvNeXt}{Colab}, \href{https://github.com/facebookresearch/ConvNeXt}{MMClassification} 
        & Reexamine the limitation of pure ConvNets and discover several key components to design ConvNets \\

        Uninet \cite{liu2022uninet} & ImageNet-1K/21k & \href{https://github.com/sense-x/uninet}{PyTorch} & Study the learnable combination of convolution, transformer, and MLP with a new search approach \\
        
        CBAM \cite{woo2018cbam} & ImageNet-1K &  \href{https://github.com/luuuyi/CBAM.PyTorch}{PyTorch} & Introduce spatial and channel attention into residual block \\

        Res2Net \cite{gao2019res2net} & ImageNet, CIFAR & \href{https://github.com/PaddlePaddle/PaddleClas/blob/release/static/docs/en/advanced_tutorials/distillation/distillation_en.md}{PaddlePaddle},
        \href{https://github.com/Jittor/jittor}{Jittor},
        \href{https://github.com/Res2Net/Res2Net-PretrainedModels}{PyTorch}
        & Propose a multi-scale residual feature learning module \\
        
        \hline
    \end{tabular}
\end{table*}
\begin{table*}[htbp]
    \centering
    \label{table2}
    \caption{Deep neural networks based on residual learning for object detection}
    \renewcommand{\arraystretch}{1.5} 
    \label{tab:references}
    \begin{tabular}{|l|l|l|p{5cm}|}
        \hline
        \textbf{Reference} & \textbf{Dataset} & \textbf{Code} & \textbf{Description} \\
        \hline
        
        Mask R-CNN\cite{he2017mask}  & COCO & 
        \href{https://github.com/facebookresearch/Detectron}{PyTorch} & Use ResNet as block to learn features for object detection and instance segmentation \\
        
        YOLOv3\cite{redmon2018yolov3}  & COCO & 
        \href{https://pjreddie.com/yolo/}{PyTorch} & Introduce newfangled residual network into YOLO arichecture for one stage object detection \\
        
        MaxViT \cite{tu2022maxvit} & COCO &  \href{https://github.com/google-research/maxvit}{TensorFlow}& Propose a new block attention module to learn  local and global attention in residual unit \\

        ConvNeXt \cite{liu2022convnet} & COCO & \href{eexamine the limitation of a pure ConvNets and discover several key components to design ConvNets}{PyTorch}
        & Modify the  inverted residual bottleneck block with a large kernel depthwise convolution, Layer Normalization, and GeLU activation\\
        
        ScratchDet\cite{zhu2019scratchdet}   & PASCAL VOC 2007, COCO & \href{https://github.com/KimSoybean/ScratchDet}{Caffe} & Training single-shot object detectors from scratch with ResNet as backbone \\

        PoolNet\cite{liu2019simple} & ECSSD, PASCALS, DUT-OMRON, HKU-IS, SOD & 
        \href{https://github.com/backseason/PoolNet}{PyTorch}
        & Enhancing object detection accuracy with a U-shaped architecture featuring global guidance and feature aggregation modules\\
        
        CPD \cite{kelchtermans2004cpd}& ECSSD,PASCALSDUT-OMRON, HKU-IS, SOD& 
        \href{https://github.com/wuzhe71/CPD}{PyTorch}
        & Transforming the problem of aligning two point sets into a probability density estimation problem \\
        
        OPANAS \cite{liang2021opanas}  &PASCAL VOC 2007, MS COCO & \href{Investigating Multiple Information Path Aggregation Realization Searches for Faster and Stronger Feature Pyramid Structures}{PyTorch} & Investigating Multiple Information Path Aggregation Realization Searches for Faster and Stronger Feature Pyramid Structures \\
        
        RetinaNet \cite{lin2017focal}& COCO & \href{https://github.com/yhenon/pytorch-retinanet}{PyTorch} & Propose a new loss function (Focal loss) aiming to  focus on hard samples on ResNet-FPN architecture  \\
        
        $U^2$-Net\cite{qin2020u2} & COCO & \href{https://github.com/yhenon/pytorch-retinanet}{XXX} & Proposed Residual U-blocks which nested  UNet-like unit into residual block\\ 
        
        \hline
    \end{tabular}
\end{table*}
\begin{table*}[htbp]
    \centering
    \label{table3}
    \caption{Deep neural networks based on residual learning for segmentation}
    \renewcommand{\arraystretch}{1.5} 
    \label{tab:references}
    \begin{tabular}{|l|l|l|p{5cm}|}
        \hline
        \textbf{Reference} & \textbf{Dataset} & \textbf{Code} & \textbf{Description} \\
        \hline
        
        FCN\cite{long2015fully} & PASCAL VOC 2011 & \href{https://github.com/shekkizh/FCN.tensorflow}{TensorFlow}
        & Use the skip connection to  combine semantic information \\

        
        U-Net \cite{ronneberger2015u} &VNC/PhC-U373/DIC HeLa & 
        \href{https://lmb.informatik.uni-freiburg.de/people/ronneber/u-net/}{Caffe}
         & Skip connections  connect feature maps from the encoder to corresponding feature maps in the decoder \\

        SegNet \cite{badrinarayanan2017segnet} &CamVid/SUN &
        \href{https://github.com/alexgkendall/caffe-segnet}{Caffe} & SegNet utilizes skip connection with the pooling indices from the max-pooling layers in the encoder \\
        
        DiSegNet \cite{xu2021disegnet}  & PET/CT & \href{https://github.com/Francesco-Voto/Disegnetti}{iOS} & Passing the pooled features to the decoder with skip connection\\

        Unet++ \cite{zhou2019unet++} & EM/Cell/Nuclei/Brain Tumor/Liver/Lung Nodule & 
        \href{https://github.com/MrGiovanni/UNetPlusPlus}{Keras}
        & A multi-scale feature fusion by using skip connect mechanism \\
        
        nnU-Net \cite{isensee2021nnu} & ACDC/MNMs/bRAts21 & 
        \href{https://github.com/MIC-DKFZ/nnUNet}{PyTorch}
        &  A customized UNet architecture for semantic segmentation \\
        
        ResUNet-a \cite{diakogiannis2020resunet} &ISPRS 2D Potsdam & \href{hhttps://github.com/feevos/resuneta}{MXNet} & Introduce residual blocks on top of the U-Net architecture, speeding up network convergence \\
        
        ConvNet\cite{yu2017volumetric} & MICCAI,PROMISE12 & \href{https://github.com/TorontoDeepLearning/convnet}{Gloab} & Proposes a new kind of mixed residual connections\\

        Inception-ResNet\cite{szegedy2017inception} & ImageNet & \href{https://github.com/titu1994/Inception-v4}{Keras} &  Incorporating Inception with ResNet's residuals enhances training \\

        DANet\cite{fu2019dual} & Cityscapes,PASCAL,Context,COCO Stuff& \href{https://github.com/junfu1115/DANet}{PyTorch} &Skip connections transmit features across encoder-decoder levels \\

        DeepLab\cite{chen2017deeplab}  & PASCALVOC-2012,PASCAL-Person-Part,Cityscapes & \href{https://github.com/kazuto1011/deeplab-pytorch}{PyTorch} & Skip connections maintain spatial richness, transmitting high-resolution features to the decoder\\
        
       DeepLabv3 \cite{chen2017rethinking} & PASCAL VOC 2012 & \href{https://github.com/fregu856/deeplabv3}{PyTorch} & Skip connections originate from diverse encoder levels, offering varying resolution and semantic data \\

        DeepLabv3+ \cite{chen2018encoder} &PASCAL VOC 2012,Cityscapes & \href{https://github.com/tensorflow/models/tree/master/research/deeplab}{TensorFlow} &  Skip connections connect to intermediate layers of the decoder, allowing it to leverage feature maps from different levels of the encoder \\

        ESPNet \cite{mehta2018espnet} & Cityscapes,PASCAL VOC 2012 & \href{https://github.com/sacmehta/ESPNet}{PyTorch}  & Skip connections directly link the low-resolution feature maps from the encoder to the high-resolution feature maps in the decoder \\

        ESPNetv2 \cite{mehta2019espnetv2} & Cityscapes,PASCAL VOC 2012 & \href{https://github.com/sacmehta/ESPNetv2-COREML}{PyTorch}
        & Lightweight architecture for semantic segmentation \\

        DCU-Net \cite{yang2022dcu} & DRIVE,CHASEDB1 &  \href{https://github.com/hwding-whu/DCU-Net}{Keras}
        & Enhances U-Net with skip connections for multi-level information capture \\
        
        CCNet \cite{huang2019ccnet} & Cityscapes,ADE20K &  \href{https://github.com/speedinghzl/CCNet}{PyTorch} & CCNet's skip connections merge encoder and decoder feature maps for enriched decoder context \\

        EMANet \cite{li2019expectation} & PASCAL VOC/Context,COCO Stuff & 
        \href{https://github.com/XiaLiPKU/EMANet}{PyTorch}
        & Sum the output with original input, to form a residual-like block \\

        Swin Transformer \cite{liu2021swin} & COCO,ADE20K & 
        \href{https://github.com/microsoft/Swin-Transformer}{PyTorch}
        & In the process of W-MSA and MLP, skip connections are employed \\

        TransUNet \cite{chen2021transunet} & Synap,ACDC & 
        \href{https://github.com/Beckschen/TransUNet}{PyTorch}
        & Utilizing UNet's skip connections grants direct access to multi-scale features for transmission from encoder to decoder\\

        TransGAN \cite{jiang2021transgan} & CIFAR-,STL10,CelebA & 
        \href{https://github.com/VITA-Group/TransGAN}{PyTorch}
        & Introduces the generative adversarial network architecture using self-attention mechanism capable of generating high-quality images \\

        SAM \cite{kirillov2023segment} & SA-1B,MIAP & 
\href{https://github.com/facebookresearch/segment-anything}{PyTorch}
        & Networks capable of interactive segmentation \\

        \hline
    \end{tabular}
\end{table*}

\begin{table*}[htbp]
    \centering
    \label{table4}
    \caption{Deep neural networks based on residual learning for image reconstruction, including image compression, denoise, deblur, and super-resolution}
    \renewcommand{\arraystretch}{1.5} 
    \label{tab:references}
    \begin{tabular}{|l|l|p{3cm}|p{1.2cm}|p{7cm}|}
        \hline
        \textbf{Mandate}&\textbf{Reference} & \textbf{Dataset} & \textbf{Code} & \textbf{Description} \\
        \hline
        
        &SlimCAEs\cite{yang2021slimmable} & CLIC,Kodak & \href{https://github.com/FireFYF/SlimCAE}{TensorFlow} & Rate and distortion are co-optimized for different capacities with residual connections \\

        &MAXIM\cite{tu2022maxim} & GoPro,HIDE,RealBlur-J/R,SIDD & \href{https://github.com/google-research/maxim}{PyTorch} & Remote interaction by space gated MLP via jumper connection\\
        
       Compression& GMM-and-Attention \cite{cheng2020learned} & Kodak,high-resolution & \href{https://github.com/ZhengxueCheng/Learned-Image-Compression-with-GMM-and-Attention}{TensorFlow}
         & Determination of latent code distribution parameters by discrete Gaussian mixed likelihood method and integration of attention by residual linkage.\\
        &GAN Compression\cite{li2020gan} & Cityscapes ,COCO & \href{https://github.com/mit-han-lab/gan-compression}{PyTorch}
        & Uniform unpaired and paired learning by transferring original multiple intermediate representation knowledge to its compressed model via jumper connections.\\
        \hline
        
       & D-BSN\cite{wu2020unpaired}  & ImageNet,BSD & \href{https://github.com/XHWXD/DBSN}{PyTorch} & Combining self-supervised learning and knowledge disitllation to facilitate residual learning\\

       & D2HNe\cite{zhao2022d2hnet}& D2 & 
        \href{https://github.com/zhaoyuzhi/D2HNet}{PyTorch}
        & Connecting information in long and short exposure images with residuals \\
        
        &DDRNet \cite{yan2018DDRNet} & Face Warehouse, Biwi Kinect face & 
        \href{https://github.com/neycyanshi/DDRNet}{Tensorflow}
        & Residual co-training uses fused frames and high-quality images to resolve noise \\
        
        Denoise&R2R\cite{pang2021recorrupted}&BSD68& \href{https://github.com/PangTongyao/Recorrupted-to-Recorrupted-Unsupervised-Deep-Learning-for-Image-Denoising}{PyTorch} & Unsupervised data enhancement techniques to address overfitting due to lack of real images. \\
        
        &MWDCNN\cite{tian2023multi} & BSD68,CBSD68,Kodak24 & \href{https://github.com/hellloxiaotian/MWDCNN}{PyTorch} & Optimizing the obtained features with residuals to improve denoising results \\

       & DeamNet\cite{ren2021adaptive} &AWGN,Real-world& \href{https://github.com/chaoren88/DeamNet}{PyTorch} & Residuals combined with nonlinear filters, reliability matrices, and eigentransforms to propose an adaptive prior\\

      &  NBNet\cite{cheng2021nbnet} & SIDD,DND& \href{https://github.com/MegEngine/NBNet}{PyTorch} & 
Networks combine residuals to learn signal-to-noise separation via a feature space basis \\
        \hline
      & DeFMO\cite{rozumnyi2021defmo}& Falling Objects,TbD-3D ,TbD & \href{https://github.com/rozumden/DeFMO?tab=readme-ov-file}{PyTorch} & 
Deblurring image enhances clarity, removing background noise \\

      &  PSD\cite{chen2021psd} & OTS,URHI & \href{https://github.com/zychen-ustc/PSD-Principled-Synthetic-to-Real-Dehazing-Guided-by-Physical-Priors}{PyTorch} & 
Residual Unlabeled Smoke Images Fine-tune Pre-training Backbone When Unsupervised\\

       & MIMO-UNet\cite{cho2021rethinking}& GoPro,RealBlur & \href{https://github.com/chosj95/MIMO-UNet}{PyTorch}  &Implementing a coarse-to-fine strategy for multiple inputs and outputs with jumper connections \\

     Deblur&  DarkDeblurNet\cite{sharif2023darkdeblur}& DarkShake& \href{https://github.com/sharif-apu/DarkDeblur}{PyTorch} 
        & A single image deblurring is performed under low light conditions \\

      &  MSFS-Net\cite{zhang2023multi}& HIDE,GoPro,RealBlur & \href{https://github.com/LiQiang0307/MSFS-Net}{PyTorch} &Frequency separation captures image information at different scales through residual linkage \\
        
       & UFPNet\cite{fang2023self}& GoPro,RealBlur-J &  \href{https://github.com/Fangzhenxuan/UFPDeblur?tab=readme-ov-file}{PyTorch} &Representation of kinematic fuzzy kernel fields in potential space by normalized flows \\

        &JCD\cite{zhong2021towards}& BS-RSCD & 
        \href{https://github.com/zzh-tech/RSCD}{PyTorch}
        & Fusion of features from bi-directional warping and deblurring streams using residual concatenation \\
        \hline
       & MIRNet \cite{zamir2020learning}&SIDD,DND,LOL,Adobe-MIT FiveK &  \href{https://github.com/swz30/MIRNet}{PyTorch} & Multi-scale residual block maintains precise high-res representations\\

      &  MemNet\cite{tai2017memnet} &BSD,Classic5,LIVE1 &  \href{https://github.com/tyshiwo/MemNet}{PyTorch} & Recursive and Gate Units Explicitly Explore Persistent Memory via Residual Connections\\
        
     Super-Resolution&  Idinvert\cite{zhu2020domain} &FFHQ,LSUN Tower,LSUN Bedroom  &  \href{https://github.com/genforce/idinvert_pytorch}{PyTorch} &Domain-aware GAN reverses to reconstruct input images, reversing semantically meaningful edits\\
        
       & IRN \cite{xiao2020invertible}&Set5/14,BSD100,Urban100&  \href{https://github.com/pkuxmq/Invertible-Image-Rescaling}{PyTorch} & Residual linking models the scaling down and scaling up process to capture missing information\\
        
       & RIAD\cite{zavrtanik2021reconstruction} &RIAD,MVTec AD &  \href{https://github.com/plutoyuxie/Reconstruction-by-inpainting-for-visual-anomaly-detection}{PyTorch} & Randomly remove partial image regions and reconstruct the image from partial inpaintings\\

       & ELAN\cite{zhang2022efficient} &Set5/14,BSD100,Urban100&  \href{https://github.com/xindongzhang/ELAN}{PyTorch} & Efficient extraction of local structural information in images using displacement convolution and residual joining\\
        
       & SwinIR\cite{liang2021swinir} &DIV2K,Flickr2K,OST ,WED, FFHQ,Manga109 &  \href{https://github.com/JingyunLiang/SwinIR}{PyTorch} & \textbf{A robust baseline model based on the Swin Transformer, for image restoration}\\
        
        &Uformer\cite{wang2022uformer} &GoPro,HIDE,RealBlur-J,RealBlur-R &  \href{https://github.com/ZhendongWang6/Uformer}{PyTorch} & Three jump connections are proposed to pass the output of the Transformer from the encoder section to the decoder section\\

       & RFDN\cite{liu2020residual} &DIV2K,Set5/14, BSD100,Urban100, Manga109 &  \href{https://github.com/njulj/RFDN}{PyTorch} & Multiple feature distillations use skip connections to learn more discriminative feature representations\\
        
        \hline
    \end{tabular}
\end{table*}
\section{Discussion and future works}
\label{sec7}
The main idea behind skip connection is to facilitate residual learning, making it easier to optimize the deep neural network and achieve more effective discriminative features to improve performance. In this section, we discuss the concept of skip connection in specific tasks and summarize current research trends for future directions.

\subsubsection{Skip connection in large vision models} 

With the success of large language models (LLMs), like GPTs \cite{radford2018improving}\cite{radford2019language}\cite{brown2020language}, LLaMA \cite{touvron2023llama}, and Alpaca \cite{taori2023alpaca}, it is now possible to develop a large vision model that  can handle general vision tasks. Several recent methods, including Swin Transformer \cite{liu2021swin}, DETR\cite{carion2020end}, CLIP \cite{radford2021learning}, and SAM \cite{kirillov2023segment}, aim to bridge the gap between the general vision tasks. Training a large and general vision model, which typically includes over 1 billion parameters (such as  GPT-3, which has more than 175 billion parameters), presents several challenges. One of these challenges is how to train such a model efficiently and effectively. One technique that can be used to build large vision models is residual learning, which involves introducing skip connections. Specifically, the pre-trained LLM can be utilized to generate features and design a lightweight network for learning residual features in vision tasks. This approach allows for efficient fine-tuning the LLM and training of the lightweight sub-network. We hope the large vision model developed based on residual feature learning could learn more general and discriminative features for various vision tasks.
\subsubsection{Residual learning for generative models}
Generative adversarial networks (GANs) have been a popular research topic \cite{gui2021review} since the pioneering work in \cite{creswell2018generative}. GANs offer an alternative to maximizing likelihood with the help of a discriminator and generator. However, training an unregularized GAN does not always converge \cite{mescheder2018training}. In \cite{gnanha2022residual}, the authors proposed a residual generator for GANs, which is robust against model collapse and can improve the generation quality of GANs. In theory, the goal of training a generative model is to mimic the distribution of a real dataset. Therefore, it is possible to update the generative model in a residual way by computing the residual distribution for each iteration. 
A recent type of generative model, the denoising  diffusion model \cite{ho2020denoising}\cite{nichol2021improved}, has demonstrated its effectiveness in generating images. This model consists of two stages: a forward diffusion stage and a reverse diffusion stage. During the forward diffusion stage, Gaussian noise is gradually added to the input data. During the reverse diffusion stage, the model attempted to recover the original input data by learning the reversed diffusion process \cite{croitoru2023diffusion}. One of the limitations of diffusion models is their slow speed due to  the high number of iterative steps required during sampling. A potential future direction is to introduce residual learning, using skip connections, into diffusion models to accelerate the running speed.
\subsubsection{Residual correction in reinforcement learning}
Following the success of AlphaGo \cite{silver2016mastering} and AlphaZero \cite{silver2018general}, reinforcement learning (RL) has demonstrated a significant improvement in interactive environments, such as gaming and self-control in robots. The key to reinforcement learning is to maximize some notion of cumulative reward. However, it is not trivial to learn the best policy to respond to the environment for the next action. To update the value and policy networks, sampling in a large space is typically required. In \cite{johannink2019residual}, residual reinforcement learning is proposed for robot control, where the control problem is divided into two parts: feedback from the conventional control methods and residual solved with RL. In \cite{silver2018residual}, residual policy learning is used to improve performance in complex robotic manipulation problems. It has been shown that incorporating residual correction on top of the initial controller can result in significant improvements. Therefore, it may be beneficial to consider implementing residual correction in RL and other tasks, such as training a large and general vision model, in the future.
\subsubsection{Image reconstruction with skip connection in deep neural network} 
Image reconstruction is a classical topic and ongoing research topic that encompasses signal sampling \cite{candes2008introduction}, noise reduction \cite{mallat1999wavelet}, and high-resolution reconstruction \cite{chan2003wavelet}. Significant progress has been made in image reconstruction using deep neural networks over the past two decades. However, deep neural networks have a limitation in that they require a large amount of data for training and corresponding ground truth for supervised or semi-supervised learning.  Some tasks, such as image denoising and restoration, are not only time-consuming but also labor-intensive. Particularly, it is likely impossible to obtain the original images and the corresponding labels in some tasks, such as image denoising and image restoration. With the development of large generative image models and large general vision models, extracting general features may be helpful for these models.  A specific module can be designed to learn residual features for fine-tuning the inputting general features for image reconstruction. It is speculated that the following research directions will receive more attention in the future.

\section{Conclusion}
This paper surveys the history of skip connections and residual learning. It summarizes the development of skip connection in deep neural networks, mainly including how to improve discriminative feature learning and trade-off the accuracy and efficiency with residual learning. The effectiveness of skip connection in residual learning is explained. Finally, we discussed some of the open questions and promising directions for using skip connections and residual learning in deep neural networks. We hope this survey will inspire peer researchers to work on this topic in the future.

\section*{Acknowledgements}
This work is supported by the Guangdong Provincial Key Laboratory of Human Digital Twin (No. 2022B1212010004), the Hubei Key Laboratory of Intelligent Robot in Wuhan Institute of Technology (No. HBIRL202202), the Open Project Program of the State Key Laboratory of CAD and CG in Zhejiang University (No. A2322), and the Fundamental Research Funds for the Central Universities of China (No. PA2023IISL0095), the 15th Graduate Education Innovation Fund of Wuhan Institute of Technology (No. CX2023319).

\section*{Conflict of Interest Statement}
The authors have no relevant conflicts of interest to disclose.



 
%

\vfill

\end{document}